\documentclass{egpubl}
\usepackage{pg2023}

\SpecialIssueSubmission    

\CGFStandardLicense

\usepackage[T1]{fontenc}
\usepackage{dfadobe}  

\usepackage{cite}  
\BibtexOrBiblatex
\electronicVersion
\PrintedOrElectronic
\ifpdf \usepackage[pdftex]{graphicx} \pdfcompresslevel=9
\else \usepackage[dvips]{graphicx} \fi

\usepackage{egweblnk} 

\usepackage{amsmath}
\usepackage{amsfonts}
\usepackage{amssymb}
\usepackage{bm}
\usepackage{tabularx}
\usepackage{xcolor}
\usepackage{pifont}
\usepackage{colortbl}
\usepackage[export]{adjustbox}

\newcommand{\R}{\mathbb{R}}

\newcommand{\dd}{\text{d}}

\title[Transfer Function Optimization for Comparative Volume Rendering]%
      {Transfer Function Optimization for Comparative Volume Rendering}

 \author[C. Neuhauser \& R. Westermann]
 {\parbox{\textwidth}{\centering Christoph Neuhauser$^{1}$\orcid{0000-0002-0290-1991} and Rüdiger Westermann$^{1}$\orcid{0000-0002-3394-0731}
         }
         \\
 {\parbox{\textwidth}{\centering $^1$Technical University of Munich, School of Computation, Information and Technology, Germany
        }
 }
}

\begin{document}

\teaser{
\vspace{-0.6cm}
\begin{tabularx}{1.0\linewidth}{X<{\centering}X<{\centering}X<{\centering}}
(a) Mutual information & (b) Pearson correlation & (c) \textit{(b)} with optimized transfer function \\
 \includegraphics[width=0.9\linewidth,valign=m]{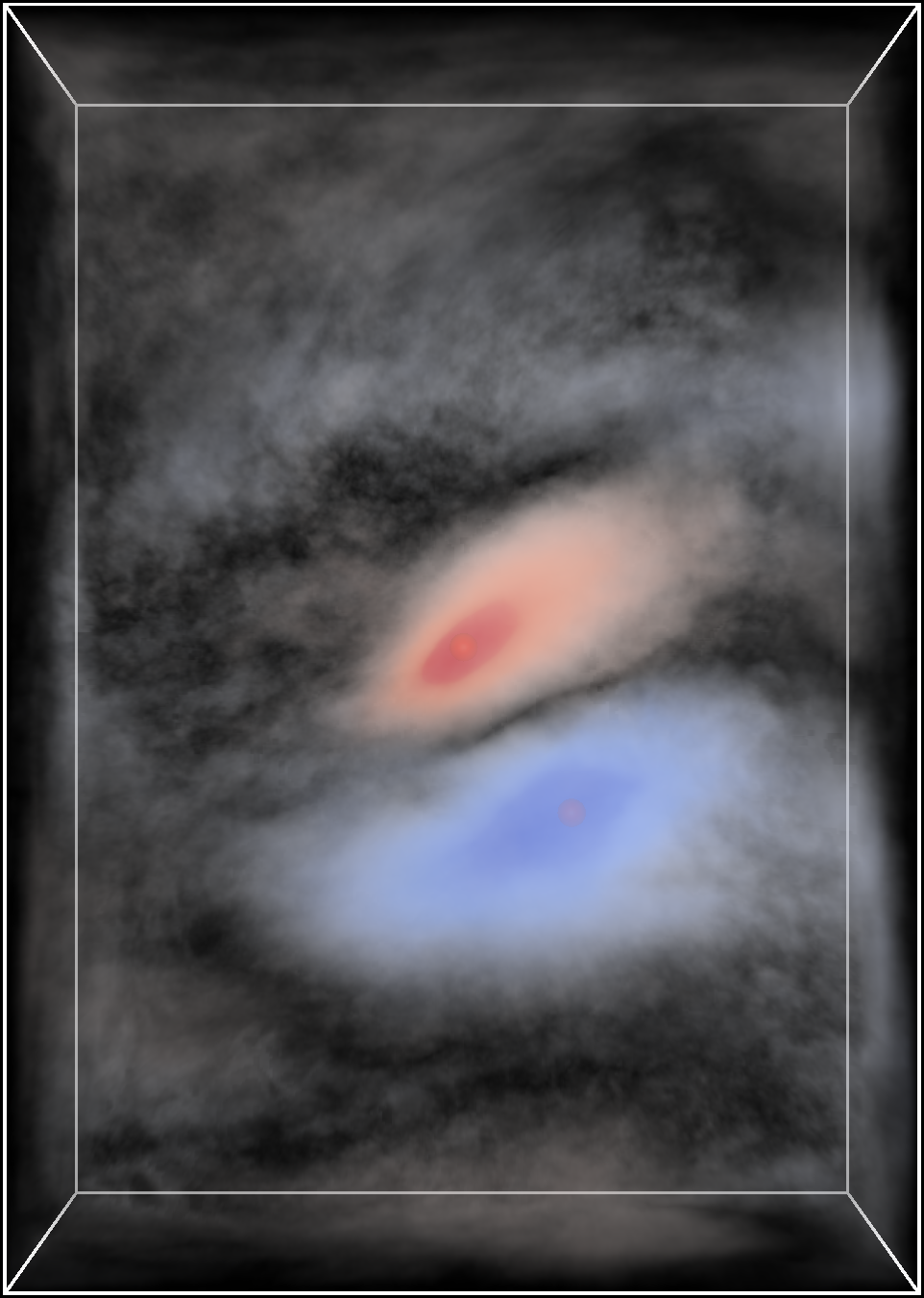} &
 \includegraphics[width=0.9\linewidth,valign=m]{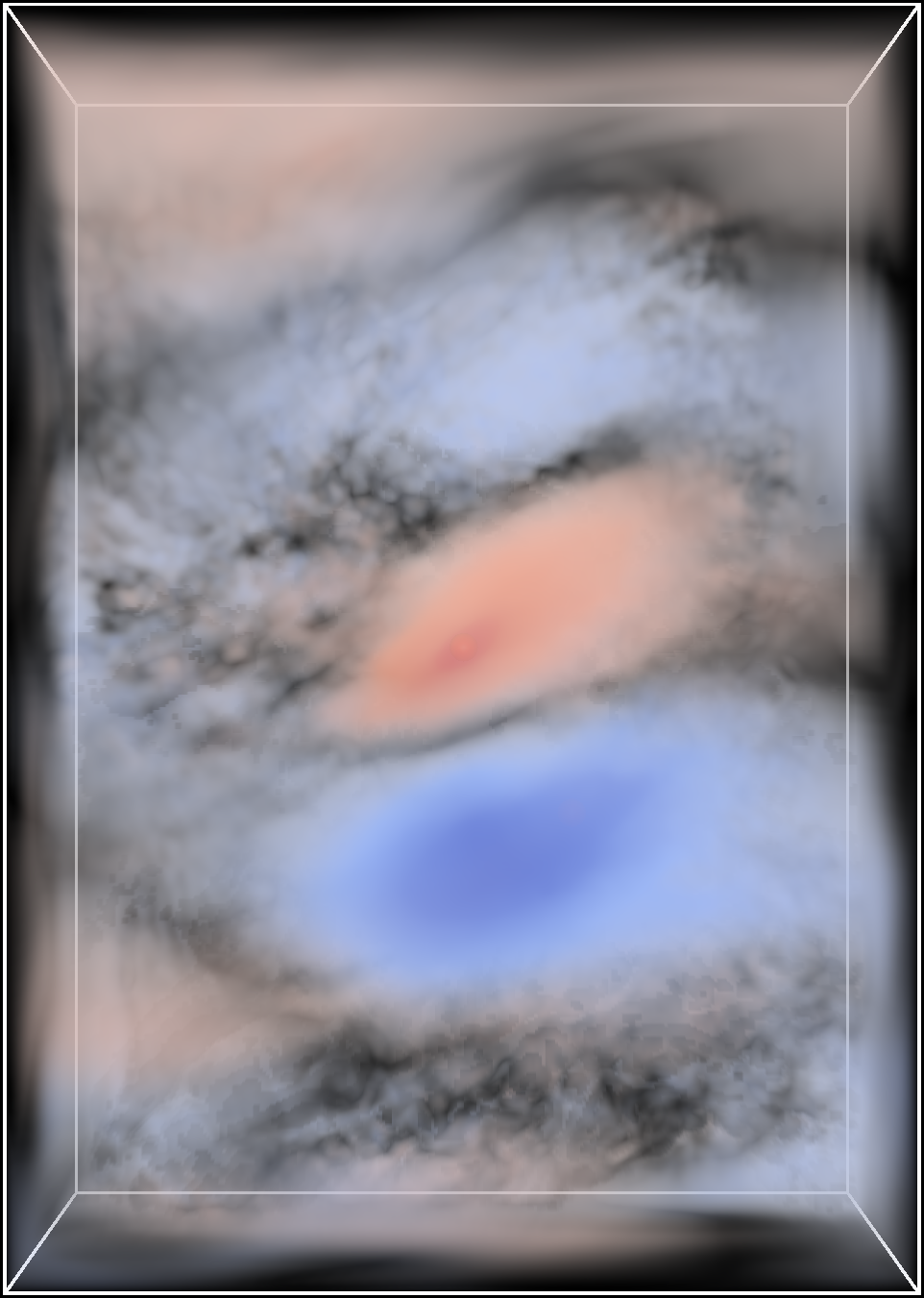} &  \includegraphics[width=0.9\linewidth,valign=m]{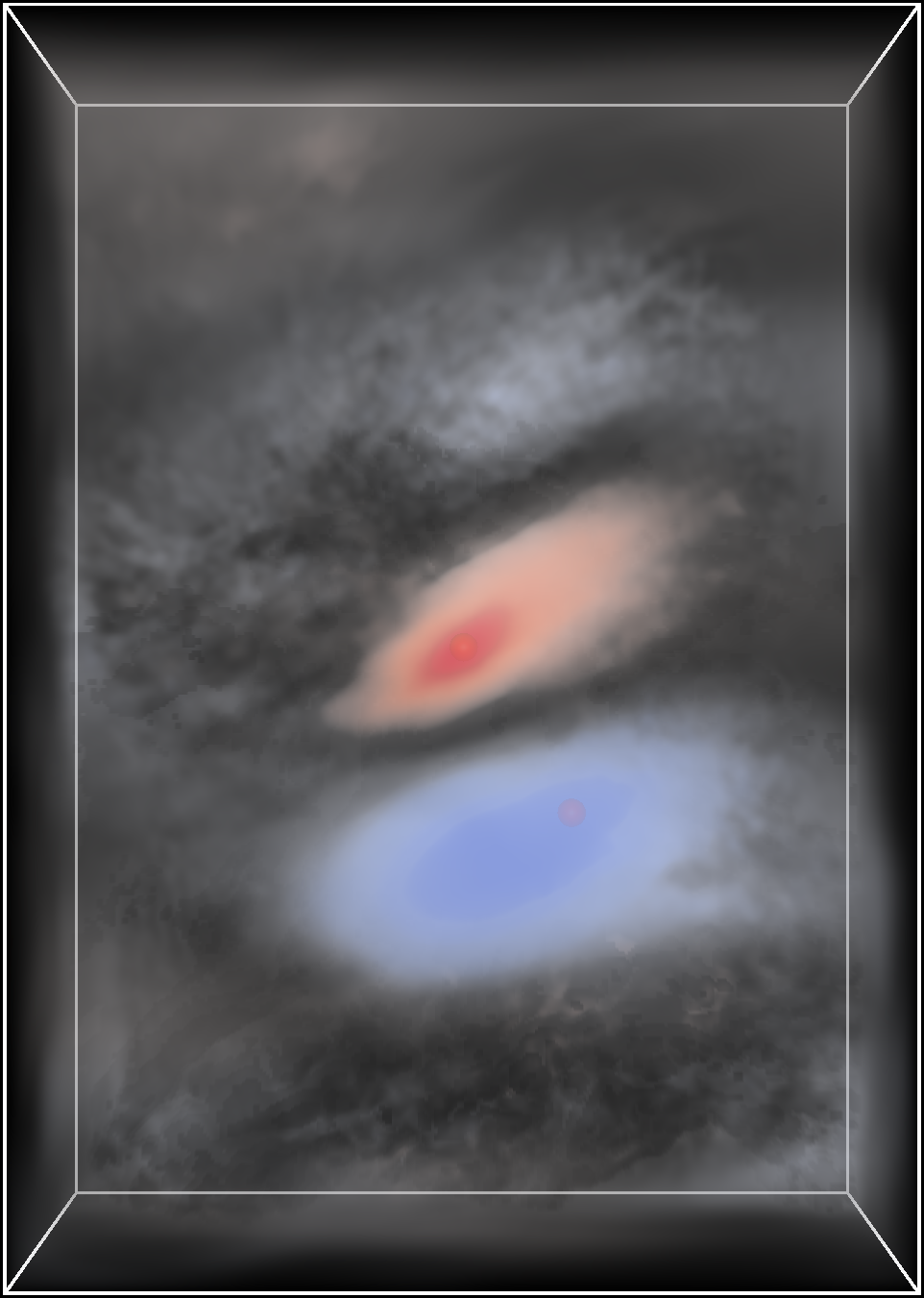}  \\
\end{tabularx}
    \caption{Left: Volume rendering of mutual information in a simulation ensemble, using a manually selected color and opacity transfer function. Middle: For the same ensemble, the corresponding Pearson correlation field is rendered using the same transfer function. Right: The Pearson correlation field is rendered using a transfer function that has been optimized to generate images as similar as possible to those showing mutual information.
    By rendering a 3D residual map of pre-shaded voxel colors, structural differences in the scalar fields can be further conveyed.  
    }
    \label{fig:teaser}
}

\maketitle
\begin{abstract}
   Direct volume rendering is often used to compare different 3D scalar fields. The choice of the transfer function which maps scalar values to color and opacity plays a critical role in this task. We present a technique for the automatic optimization of a transfer function so that rendered images of a second field match as good as possible images of a field that has been rendered with some other transfer function. This enables users to see whether differences in the visualizations can be solely attributed to the choice of transfer function or remain after optimization. We propose and compare two different approaches to solve this problem, a voxel-based solution solving a least squares problem, and an image-based solution using differentiable volume rendering for optimization. We further propose a residual-based visualization to emphasize the differences in information content.


\end{abstract}

\section{Introduction}

In direct volume rendering of a 3D scalar field using an emission-absorption optical model, the scalar values are first mapped to color and opacity via a so-called transfer function. Such a transfer function is a mapping $T: [min, max] \rightarrow [0, 1]^4$, which is commonly realized via a table with $N$ entries, each containing $RGB\alpha$ values. Here, $\alpha$ refers to opacity, and scalar values are transformed so that the data is linearly mapped to table entries $0$ to $n_T-1$. 


The choice of the transfer function is important for emphasizing interesting structures via high opacity and specific colors, at the same time suppressing non-relevant structures by making them transparent. There are many different approaches for designing transfer functions for a given field, many of which have been reviewed in the report by Ljung et al.~\cite{TrafoDVR}. Transfer functions can be designed either manually by the user via a transfer function editor~\cite{Kindlmann}, 
or automatically via the mapping of derived importance measures like derivatives \cite{Levoy-gradientshading,MultidimTrafo} or mathematical optimization with respect to different objective functions \cite{FeatureOpacity,VoxelVisibilityModel,TFInfoDivergence,DiffDVR}. 


The design of a transfer function for a single field can be tedious, and this process becomes even more cumbersome when a transfer function needs to be designed that makes a second field look as similar as possible to a reference field. This task is required in applications where volume rendering is used to compare two or more volumetric fields (see Fig.~\ref{fig:teaser}). In such applications, using the same transfer function as for the reference data can give significantly different renderings, even though the two fields may have very similar structure. For instance, this is the case when the second data set is just a scaled copy of the reference.    
If it is possible to choose a transfer function that minimizes the visual differences between the two fields, it becomes possible to see in which regions these differences cannot be explained by the choice of transfer function and, thus, manifest in the structure of the data.

\subsection{Contribution}

In this work, we address the problem of computing such an optimized transfer function automatically. We propose two different approaches, which solve the optimization problem in voxel- or image-space. 

The first approach aims to minimize the difference in voxel colors and opacities after applying the transfer functions to the scalar voxel values. We formulate this task as a least squares problem, and shed light on different strategies for computing a solution. We present different solver implementations and evaluate their advantages and disadvantages.

The second approach uses image-based optimization via differentiable direct volume rendering (DiffDVR). In a recent work by Weiss and Westermann \cite{DiffDVR}, DiffDVR has been used to optimize a transfer function for a single field, so that reference images of this field with unknown transfer function can be reproduced by rendering the field using the optimized transfer function. We build upon this approach to optimize for a transfer function that matches images of the reference field. 

Both optimization approaches are evaluated via use cases from numerical fluid simulation. We demonstrate the application for comparing different correlation measures in ensemble simulations and different physical parameters from a single simulation run. We report cases where applying the optimized transfer function results in minor visual differences, which indicate high similarity between the two fields. In other situations such a transfer function cannot be found, hinting at structural differences in the data distributions. To support a more effective comparison, we propose using a residual map, which stores voxel-wise color differences and can be rendered interactively from arbitrary perspectives. To emphasize differences that more strongly affect the visibility of different structures, the residual map stores the $l2$-norm of opacity-weighted color values. 

The proposed optimization framework is written in C++, Vulkan and CUDA. Most solvers run on both the CPUs and GPUs of any hardware vendor. GPU solvers using sparse matrix-matrix and sparse matrix-vector multiplications employ the CUDA library cuSPARSE~\cite{cuSPARSE}, and they are thus restricted to NVIDIA GPUs. We provide a graphical user interface for the direct volume renderer and probing different solvers, as well as loaders for common data formats, like NetCDF, VTK file formats and GRIB. The code is made publicly available under a BSD license at \URL{https://osf.io/s9fdx/?view_only=d02ec3fdd5e64c7a931533711d914e33}. For review purposes, the code has been anonymized.


\section{Related Work}

Especially in volume visualization, the automatic design of a transfer function has been addressed in a number of previous publications. Some of these approaches, similar in spirit to our approach, make use of optimization to address this problem. The recent survey by Sun et al.~\cite{MathOptVis} on mathematical optimization in data visualization addresses some of these approaches. 

Gradient-based optimization of a transfer function using information measures has been proposed by Ruiz et al.~\cite{TFInfoDivergence}. The approach computes voxel density histograms and uses a target histogram to quantify a current visibility distribution when viewed from different perspectives. The optimizer uses the effect of density variations on visibility to obtain a transfer function that optimizes information content in the generated images.

Chen et al.~\cite{DVRDichromats} optimize the color transfer function 
such that the generated images match the output of an image recoloring algorithm as closely as possible. They employ a least squares method on the CPU, but do not discuss the specific solver they use. In order to keep the optimization process fast, pixel subsets are sampled randomly for optimization. A drawback of the proposed technique is that it can only be used for optimizing the color transfer function, as in DVR the pixel color is linear in the sampled colors, but not in the opacities, i.e., 
$C_{pixel} = \sum_{i=1}^S C_i \alpha_i \prod_{j=1}^{i-1}(1 - \alpha_j)$.

Berger et al.~\cite{berger2018generative} use neural networks to learn the mapping from an image that is generated via a certain transfer function to a new image as it would appear when a different transfer function is applied.  
Weiss and Westermann~\cite{DiffDVR} introduce DiffDVR via automatic differentiation to optimize external parameters of the rendering process, such as the viewpoint, the transfer function, and the integration stepsize. To enable such an optimization, different loss functions are used to steer these parameters towards the optimal solution. 
We build upon this framework as one possible approach to compute an optimized transfer function for comparative volume rendering. 

\begin{figure*}[t]
    \centering
\setlength{\tabcolsep}{1pt}
\begin{tabular}{ccccc}
\includegraphics[width=1.0\linewidth]{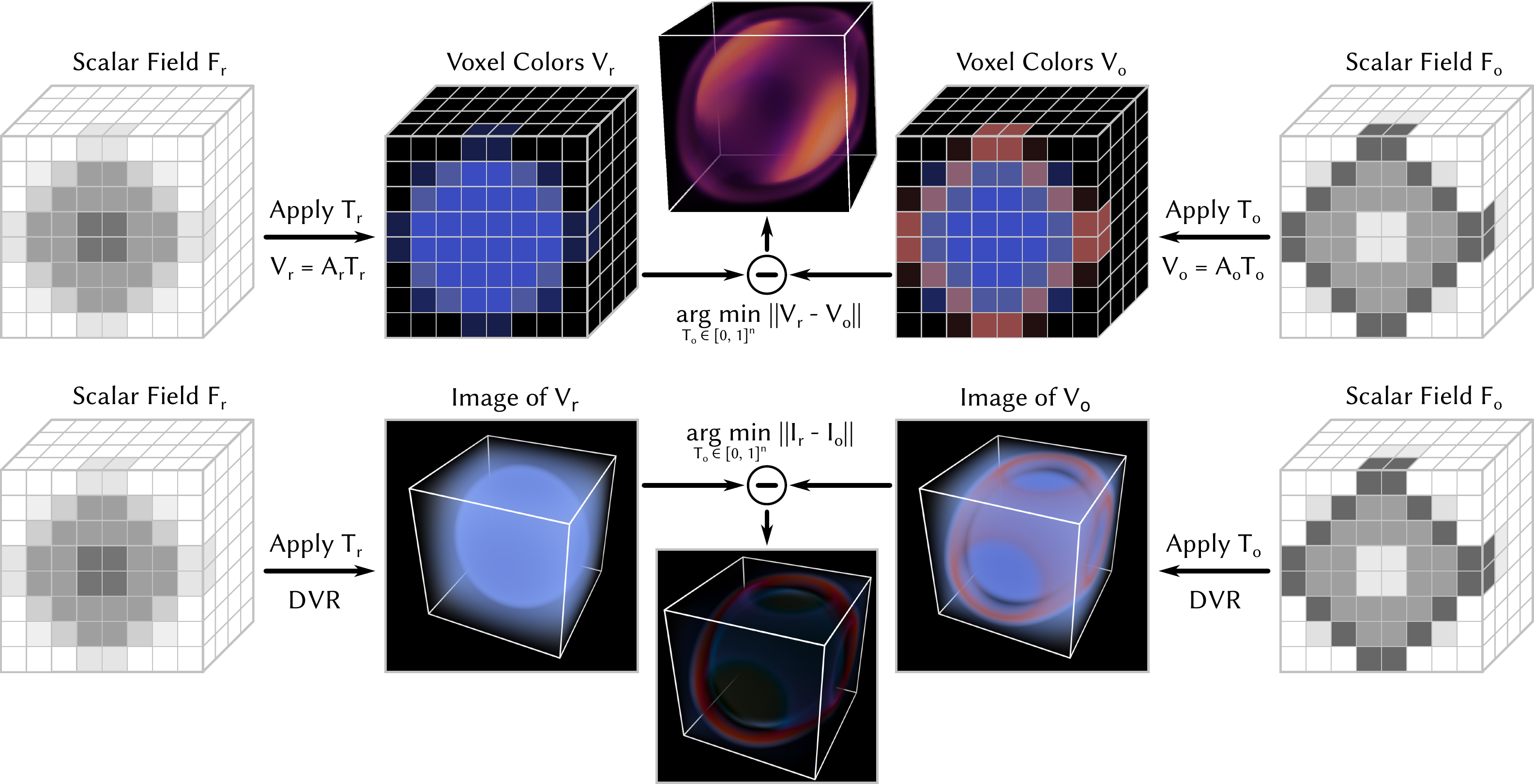}
\end{tabular}
    \caption{Proposed pipelines for transfer function optimization. $F_{\{r,o\}}$, respectively, are the reference field (with transfer function $T_r$) and the field for which the transfer function ($T_o$) is optimized. Top: Voxel-based approach using a least squares solver considering per-voxel color-residuals. $A_{\{r,o\}}$ are the system matrices relating transfer function entries to voxel colors. $V_{\{r,o\}}$ are the voxel colors. Bottom: Image-based approach using automatic differentiation of pixel values with respect to transfer function parameters. $I_{\{r,o\}}$ are the images of $F_{\{r,o\}}$ with transfer functions $T_{\{r,o\}}$.}
    \label{fig:overview}
\end{figure*}

A different use case has been addressed via opacity optimization by Günther et al.~\cite{OpacityOpt2013,OpacityOpt2014a,OpacityOpt2014b,DecoupledOpacityOpt}. This approach aims to maximize the visibility of important line structures in streamline-based flow visualization, with extensions to mesh and point data. Similar to our work, they do this by minimizing a bounded-variable least-squares problem.
Günther et al.~\cite{OpacityOpt2013} reduce the constrained least squares problem to a quadratic programming problem using the reflective Newton method~\cite{ReflectiveNewton} available via the built-in MATLAB function \texttt{quadprog}. 
In more recent work~\cite{DecoupledOpacityOpt}, Günther et al.\ have shown that by removing the spatial smoothing term in the optimization, it becomes possible to independently and analytically minimize the energy terms per pixel. Ament et al.~\cite{ExtinctionOptVols} build upon the work by Günther et al.\ and optionally employ the Levenberg-Marquardt algorithm~\cite{Levenberg} for solving non-linear least squares problems together with a preconditioned conjugate gradient method~\cite{ConjGrad} implemented in the Ceres Solver library~\cite{Ceres}. 
The authors put forward that computing the numerical solution takes a significant amount of time, restricting the problem size they can handle. Similarly, Wang et al.\ and Qin et al.~\cite{FeatureOpacity,VoxelVisibilityModel} use optimization processes for maximizing the visibility of important features in volume visualizations, thus replacing the need for manual adaption of the opacity transfer function in a visualization feedback loop. 
Läthén et al.~\cite{TFShiftVessels} optimize the transfer function shift for optimal display of the vessel structure in direct volume renderings for blood vessel visualization.

In addition to using a least squares formulation and automatic differentiation for computing an optimized transfer function, we have also considered histogram matching~\cite{HistMatching}. Histogram matching is used in image processing to modify the colors in one image so that the histogram matches a reference histogram (for contrast enhancement) or the histogram of another image. In principle, histogram matching can also be applied to scalar field histograms. However, unlike the optimization approaches utilized in this work, histogram matching cannot reverse monotonically decreasing dependencies between even the most trivial fields. For example, consider a scalar field with constant gradient into one direction, and another field with exactly the same histogram but gradients pointing into the inverse direction. In this case, histogram matching assumes that the optimal result has already been reached.

\section{Background and Overview}\label{sec:method}

We introduce a framework for the automatic optimization of a transfer function $T_{o}$ of a scalar field $F_o$ (optimization field) such that it visually matches images of a reference field $F_r$ with given transfer function $T_{r}$. Direct volume rendering is used for generating these images. 

\subsection{Direct Volume Rendering}

Let $D : \R^3 \rightarrow [0,1]$ be the scalar density volume and $r$ be a view ray through the volume.
Let $T_{\alpha}: [0,1]\rightarrow [0,1]$ be the opacity and $T_C: [0,1]\rightarrow [0, 1]^3$ the color of a density sample. 
The color reaching the eye along a ray through the volume from start point $r(a)$ to end point $r(b)$ can then be computed as 
\begin{equation}
    L(a,b) = \int_a^b g(D(r(t))) e^{-\int_a^t T_{\alpha}(D(r(u)))\dd u} \dd t.
    \label{eq:integral2}
\end{equation}
$g(d) = T_{\alpha}(d)T_C(d)$ is the emission, and the exponential term expresses the light attenuation from  the ray point $r(t)$ to the observer. 

We assume that density values are given at the vertices $\mathbf{v}_i$ of a rectangular grid, i.e., the voxels, 
so that trilinear interpolation can be used to obtain the density at arbitrary points in the domain. 
Both functions $T_{\alpha}$ and $T_C$ are discretized into $n_T$ regularly spaced bins with linear interpolation in between. 
The volume rendering integral is approximated by discretizing the ray into $S$ segments over which the opacity $\alpha_i$ and color $C_i$ are assumed constant. 
For rendering, a segment's transparency is approximated based on the Beer-Lambert model $\alpha_i=1-\exp(-\Delta t \alpha(d_i))$, where $d_i$ is the sampled volume density.
The resulting Riemann sum can be computed in front-to-back order using iterative application of alpha-blending, i.e., $L = L + (1-\alpha) L_i$, and $\alpha = \alpha+(1-\alpha)\alpha_i$.


\subsection{Optimization Methods}\label{sec:problem-formulation}

For the optimization of $T_o$ we consider two different approaches. A voxel-based approach first applies the transfer functions to pre-shade the density values at each vertex of the reference and the optimization field.
The optimization process tries to minimize the differences between the colors of all pairs of corresponding vertices in both fields, so that the generated images are as close as possible. On the other hand, an image-based approach directly minimizes the differences between the rendered images of the two fields, where during rendering the transfer functions $T_r$ and $T_o$ are considered to post-shade the sampled density values. Since an image-based loss is considered, the differences between voxels that have a larger effect on the final image are prioritized. An overview of both approaches is shown in \autoref{fig:overview}.
In the voxel-based approach, the colors that are obtained via post-shading depend linearly on the transfer function entries. I.e., the color at voxel $i$ with density $D_i$ is $V_i = (C_i, \alpha_i) = (1 - w(i)) T_{j(D_i)} + w(i) T_{j(D_i) + 1}$, where $j$ quantizes density values to bins of the transfer function and $w(i)$ is the interpolation weight of the two closest transfer function entries. The individual equations for all voxels can be reformulated as a system of linear equations $AT = V$, where $A \in \mathbb{R}^{m \times n}$ is the system matrix representing the relation between transfer function entries and voxel colors. As there is one row per $RGB\alpha$ entry, $m = 4m_V$ and $n = 4n_T$. In our case, we aim at finding a transfer function $T_o$ such that $A_o T_o$ most closely matches the colors and opacities $V_r$ of the reference volume. We will subsequently refer to $RGB\alpha$ values as colors. This linear system of equations is over-determined, as, in general, $m \gg n$, i.e., the number of voxels is considerably larger than the number of transfer function entries. In the following section, we show how to formulate the problem as a least squares problem that can be solved in an optimal way. 


In the image-based approach, we use automatic differentiation to propagate changes in the rendered images due to the variation of the transfer function back to the transfer function parameters, i.e., the colors it stores. In automatic differentiation for volume rendering \cite{Mitsuba2,DiffDVR}, the derivative of the pixel color with respect to changes in the transfer function is computed by considering the sequence of compositing operations along a view ray, and using the chain rule to obtain the final derivative from the individual derivatives of each compositing operation. Automatic differentiation can be performed either in forward mode or in reverse mode. In the former, variables are replaced by their derivatives with respect to the optimized parameters, and these derivatives are propagated in turn during ray traversal.  
In reverse mode, the operations and intermediate results are first computed and stored, and they are then traversed in reverse order to propagate the changes to the parameters to be optimized. 

Conceptually, our implementation of DiffDVR follows exactly the implementation by Weiss and Westermann, including exploitation of the ``inversion trick'' to overcome scalability limitations of the reverse mode. Since the compositing operation is invertible, reverse mode can be split into a first forward pass, which stores intermediate colors, and a second backward pass, where the order of operations is reversed and the derivatives are propagated by reusing the intermediate values. With the inversion trick, the memory requirement can be reduced significantly, as intermediate results no longer need to be stored. In every iteration, random camera poses are sampled and images for $F_r$ and $F_o$ are rendered at a resolution of $512 \times 512$ pixels in the forward pass. By traversing the samples along the view rays in the opposite order, the gradients wrt.\ the individual entries of the transfer function are computed in the backward pass.


\subsection{Residual-based Difference Visualization}

Once an optimized transfer function is computed, images of the reference field (with the initial transfer function $T_r$) and the optimization field (with optimized transfer function $T_o$) can be shown side by side. Alternatively, image-based difference measures like the root mean squared error (RMSE), the peak-signal-to-noise ratio (PSNR), the structural similarity index (SSIM)~\cite{wang2004image} or the Learned Perceptual Image Patch Similarity metric (LPIPS)~\cite{zhang2018perceptual} can be used to compare how close two images are. In this way, it can be revealed whether there are structural differences in the two fields that cannot be compensated by the optimized transfer function. 

To support an improved analysis guiding towards those regions in the domain where these differences are high, we support a DVR mode in which the residual field between the reference and the optimized field is rendered. Therefore, first the transfer functions $T_r$ and $T_o$ are applied to either field, and the residual is computed by first subtracting the per-voxel colors from the two fields and then computing, e.g., the $l2$-norm of the $RGB\alpha$ difference vector. 
We use RGB colors pre-multiplied with $\alpha$ to ensure that highly transparent regions have less contribution to the residual rendering.

\section{Voxel-based Optimization}\label{sec:least-squares}

\begin{figure}
\[
\underbrace{\begin{pmatrix}
w(1)   & 1-w(1) & 0      & 0      & 0      \\
0      & 0      & 0      & w(2)   & 1-w(2) \\
0      & 0      & w(3)   & 1-w(3) & 0      \\
0      & w(4)   & 1-w(4) & 0      & 0      \\
0      & 0      & 0      & w(5)   & 1-w(5) \\
w(6)   & 1-w(6) & 0      & 0      & 0      \\
\end{pmatrix}
\begin{pmatrix}
T_{o,1} \\ T_{o,2} \\ T_{o,3} \\ T_{o,4} \\ T_{o,5}
\end{pmatrix}}_{A x}
\]
\[
\underbrace{\begin{pmatrix}
\bullet & 0       & 0       & 0       & \bullet & 0       & \cdots  & \cdots  & 0       \\
0       & \bullet & \ddots  & \ddots  & \ddots  & \bullet & \ddots  &         & \vdots  \\
0       & \ddots  & \bullet & \ddots  & \ddots  & \ddots  & \ddots  & \ddots  & \vdots  \\
0       & \ddots  & \ddots  & \bullet & \ddots  & \ddots  & \ddots  & \bullet & 0       \\
\bullet & \ddots  & \ddots  & \ddots  & \ddots  & \ddots  & \ddots  & \ddots  & \bullet \\
0       & \bullet & \ddots  & \ddots  & \ddots  & \bullet & \ddots  & \ddots  & 0       \\
\vdots  & \ddots  & \ddots  & \ddots  & \ddots  & \ddots  & \bullet & \ddots  & \vdots  \\
\vdots  &         & \ddots  & \bullet & \ddots  & \ddots  & \ddots  & \bullet & 0       \\
0       & \cdots  & \cdots  & 0       & \bullet & 0       & 0       & 0       & \bullet \\
\end{pmatrix}
\begin{pmatrix}
T_{o,1R} \\ T_{o,1G} \\ T_{o,1B} \\ T_{o,1\alpha} \\
T_{o,2R} \\ T_{o,2G} \\ T_{o,2B} \\ T_{o,2\alpha} \\
\vdots \\
T_{o,m_TR} \\ T_{o,m_TG} \\ T_{o,m_TB} \\ T_{o,m_T\alpha}
\end{pmatrix}}_{A^T A x}
\]
\caption{Top: Depiction of the sparsity pattern of the left-hand side of $Ax = b$ for a sample volume with $m_V = 6$ voxels and a transfer function with $n_T = 5$ entries. For the sake of simplicity, individual $RGB\alpha$ entries are not listed separately. The right-hand side is a vector of voxel colors of the reference volume. Bottom: Depiction of the sparsity pattern of $A^T A$ and the normal equations. ``$\bullet$'' indicates a non-zero entry. The right-hand side is $A^T$ times a vector of voxel colors of the reference volume.
$T_o$ indicates the transfer function to be optimized. }
\label{fig:system-sparsity}
\end{figure}

In the following, we formulate the least squares problem to optimize the transfer function using the voxel-based approach. 
We then continue with the mathematical foundations before looking at different solver implementations and their strengths and weaknesses. It is clear that the least squares approach is not the only way of optimizing the transfer function. If, e.g., the $l1$-norm should be used instead of minimizing the $l2$-norm of the residual $r = b - Ax$, a least absolute deviations problem needs to be solved using different solvers.
We show in \autoref{sec:grad-desc} how gradient descent methods can be used for optimizing with respect to arbitrary norms.

\subsection{Problem Formulation}\label{sec:prob-formulation}

In \autoref{sec:problem-formulation}, we showed that the combined color and opacity $V_i = (C_i, \alpha_i)$ of the $i$-th voxel with density $D_i$ can be expressed as $V_i = (1 - w(i)) T_{j(D_i)} + w(i) T_{j(D_i) + 1}$, where $j$ quantizes density values to bins of the transfer function and $w(i)$ is the interpolation weight of the two closest transfer function entries. The voxel color can be expressed as a linear combination of two transfer function entries. The individual equations for all voxels can be reformulated as a system of linear equations $AT = V$. $V \in [0, 1]^n$ is a vector of linearized $RGB\alpha$ voxel color, $T \in [0, 1]^{m}$ is a vector of linearized $RGB\alpha$ transfer function values, and $A \in \mathbb{R}^{m \times n}$ is the system matrix representing the relation between transfer function entries and voxel colors. For $m_V$ voxels and $n_T$ transfer function entries, there are $m = 4m_V$ rows and $n = 4n_T$ columns in $A$ due to the linearization of the $RGB\alpha$ values. The resulting matrix $A$ is very sparse. In \autoref{fig:system-sparsity}~top, a simplified example is given for $m_V = 6$ and $n_T = 5$, where we assume only one color component. As can be seen, there are only two non-zero entries per row. For linearized $RGB\alpha$ values, there is an additional stride of $4$ between two non-zero row entries, as the different color channels do not interact with each other. To simplify the equations in the next subsection, we set $A = A_o$, $x = T_o$ and $b = V_r$.

\subsection{Mathematical Foundations}\label{sec:math-overview}

Ordinary least squares problems~\cite{LinearLeastSquares} are of the form
\begin{equation}
    Ax + \epsilon = b,
\end{equation}
where $\epsilon$ is an error term that should be minimized. As shown before, our least squares problem requires to solve $\min_x \lVert b - Ax \rVert_2^2$, where $r = b - Ax$ is the residual. This can be expressed as minimizing the energy term $E(x) = \lVert b - Ax \rVert_2^2 = (b - Ax)^T (b - Ax) = b^Tb - 2x^TA^Tb + x^TA^TAx$. Differentiating the energy term by $x$ and equating to zero results in $\frac{\partial E(x)}{\partial x} = 2 A^TAx - 2A^Tb = 0$. This can be reformulated as the so-called normal equations~\cite{LinearLeastSquares},
\begin{equation}\label{eq:normal-equations}
    A^TAx = A^Tb.
\end{equation}
The solution $x$ can then be obtained as $x = (A^T A)^{-1} X^Tb$. If $A$ does not have full rank, a formulation using the Moore-Penrose pseudo-inverse is available. However, due to efficiency and quality considerations, it is preferable to use a linear systems solver rather than computing the pseudo-inverse.

A major drawback when using the normal equations stems from their reduced numerical stability. The condition number of the system matrix in a linear equation system is an indicator of the numerical stability of the problem. The higher the condition number, the more instable the problem becomes. Unfortunately, in our case it holds that $cond(A^TA) = cond(A)^2$. In the process of computing $A^TA$, this results in many arithmetic addition operations and potentially very large values in the obtained matrix. In line with observations by others, also our experiments show that using the normal equations is usually the fastest alternative, yet it runs into stability issues if the matrix becomes ill-conditioned. This makes the decision which approach to use dependent on the size of the data.

In search for the most suitable solver, we further have to consider that when optimizing $RGB\alpha$ values a box-constrained least squares problem needs to be solved, i.e., $\forall i \in [1, \dots, n]: 0 \leq x_i \leq 1$. For instance, 
opacity optimization approaches~\cite{OpacityOpt2013,ExtinctionOptVols} have directly formulated the constraint problem as a more general quadratic programming problem, as there are more (and often better tested) implementations of solvers for such constraint problems. Quadratic programming aims to solve 
\begin{equation}\label{eq:quadratic-programming}
    \min_x \frac{1}{2}x^T Q x + c^T x,
\end{equation}
which is equivalent to solving an ordinary least squares problem with $Q = A^TA$ and $c = -A^Tb$. This formulation, however, shares the same drawback as solving the normal equations, as the condition number of $Q$ is the quadratic condition number of $A$. Consequentially, it would be preferable to avoid evaluating $A^T A$ for very large, ill-conditioned systems.

As a reference for future sections, we would like to note that the ordinary least squares problem is convex, as the Hessian (the second-order derivative) $H(x) = \frac{\partial^2 E(x)}{\partial x^T \partial x} = 2 A^TA$ of the energy term is positive semi-definite. Positive semi-definiteness is fulfilled, as it holds for all $z \in \mathbb{R}^n$ that
\begin{equation}\label{eq:convexity}
    z^T H(x) z = 2 z^T A^T A z = 2 (Az)^T (Az) = 2 \lVert Az \rVert_2^2 \ge 0.
\end{equation}

\subsection{Implementation and Solvers}\label{sec:solvers}

Both the matrix $A$ and the matrix $A^TA$ are sparse. As was shown in \autoref{sec:prob-formulation}, only two values in a row of $A$ contain non-zero values. These values are always exactly four columns apart, since the transfer function stores one $RGB\alpha$ value per entry. Furthermore, when creating the matrix $A$ in a sparse representation format, it is more straightforward to obtain a format that stores the rows sequentially compared to the columns, as we can easily query the non-zero entries for the $i$-th voxel (i.e., row), but not the non-zero entries for the $j$-th transfer function entry (i.e., column). Consequentially, we use the compressed sparse row (CSR) format~\cite{CSRFormat} for storing $A$ when a sparse representation is needed. $A^T A$, on the other hand, is a band matrix with non-zero entries only on the diagonal, and the two diagonals four rows/columns away from the main diagonal. The distance of four is due to the separate handling of the $RGB\alpha$ channels. A graphical depiction of the sparsity pattern in $A^T A$ and the normal equations is shown in \autoref{fig:system-sparsity}~bottom.

The matrix $A \in \mathbb{R}^{m \times n}$ is tall and skinny, as, in general, $m \gg n$. As the number of voxels $m_V$ can become very large, it is not advisable to store $A$ in a dense format. For instance, for one of our---moderately sized---test data sets 
with $1.76 \cdot 10^6$ voxels, this results in almost $7$ GiB of memory for the system matrix when using 32-bit floating point precision. $A^T A \in \mathbb{R}^{n \times n}$, on the other hand, is usually quite small and can be stored in both dense and sparse format.

\subsubsection{Normal Equations-based Solvers}\label{sec:normal-eq-solvers}

The most straightforward solution is to use a normal equations-based solver. For this, we first need to compute the matrix $A^T A$ and the vector $A^T b$. Consequentially, a fast method is needed for computing the sparse matrix-matrix product $A^T A$ and the sparse matrix-vector product $A^T b$. We support this via the linear algebra library Eigen on the CPU~\cite{Eigen}, cuSPARSE on NVIDIA GPUs~\cite{cuSPARSE} and using custom matrix-free Vulkan compute code~\cite{VulkanSpec}. The matrix-free Vulkan compute code iterates over all voxels and computes the two column indices $j(D_i)$ and $j(D_i)+1$ with non-zero entries and the interpolation weight $w(i)$ from \autoref{sec:problem-formulation}. Then, it atomically adds $(1 - w(i))^2$ to the entry $(j(D_i), j(D_i))$ of $A^T A$, $w(i)^2$ to $(j(D_i)+1, j(D_i)+1)$ and $(1 - w(i))w(i)$ to $(j(D_i)+1, j(D_i))$. Due to the symmetry of the matrix, only the upper triangular part needs to be stored. Similarly, one can also compute $A^T b$. While in theory, write conflicts on the GPU when using atomic additions from multiple threads to the same memory location might slow down the execution, in practice this proved to be one of the fastest solutions to compute $A^T A$ and $A^T b$.

When $A^T A$ and $A^T b$ have been assembled, there are multiple possible ways to proceed. As $A^T A$ is relatively small, the performance differences of the used solvers are negligible. One possibility is to solve the ordinary least squares problem with a dense or sparse linear least squares solver. We have tested multiple dense solvers (QR, LLT, LDLT, $\dots$) and generally found almost no differences, apart from the fact that solvers like QR may need to use pivoting for the sake of numerical stability, as $A^T A$ is often nearly rank deficient, i.e., certain entries may be significantly smaller than others. This happens when a transfer function value is hardly referenced by any of the voxel values. The matrix can even become rank deficient if no values fall within a certain range of the transfer function.

A disadvantage of the ordinary least squares solvers is that they do not take into account the box constraint $0 \leq x_i \leq 1$. On the other hand, in our experiments these constraints were fulfilled even without enforcing them explicitly. As the ordinary least squares problem is convex (cf.~\autoref{eq:convexity}), the computed solution is also a global optimum for the constrained problem if the constraints are fulfilled. In all other cases, we use the truncation $x_i' = \max\{\min \{ x_i, 1 \}, 0\}$.

If the constraints need to be enforced explicitly, a constrained quadratic programming solver can be used (cf.~\autoref{eq:quadratic-programming}).
We support the alternating direction method of multipliers (ADMM), first introduced by Boyd et al.~\cite{ADMM} for solving convex optimization problems, via the solver library OSQP~\cite{osqp}. This solver, in particular, can exploit the sparsity present in the problem definition. A GPU implementation of OSQP is available~\cite{osqp-gpu}, yet due to the relatively small size of the matrix resulting from the normal equations we did not encounter significant speed-ups compared to the used CPU implementation, 

\subsubsection{Ordinary Least Squares Solvers}

As noted in \autoref{sec:math-overview}, a major disadvantage of the normal equations is that they are ill-conditioned due to the quadratic condition number of the underlying matrix. Therefore, we have also included the sparse linear least squares solver CGLS~\cite{LSQR} in our evaluations, including Jacobi preconditioning to achieve faster convergence. For moderately sized problems it gives results equivalent to those obtained with the normal equations, and numerically more stable results for very large or almost rank deficient problems. On the other hand, this class of solvers is in general slower than computing the matrix $A^T A$ and the vector $A^T b$.

\subsubsection{Gradient Descent Solvers}\label{sec:grad-desc}

We have also implemented a gradient descent solver supporting either a constant update rate $\alpha$ or the Adam optimizer~\cite{Adam}. Like CGLS, the gradient descent solver shows stable behaviour for more ill-conditioned systems. The solver has been implemented in a matrix-free manner by accumulating the gradients via atomic additions. Notably, the gradient descent solver supports not only solving for the optimal transfer function in the least squares sense via the $l2$-norm, but also in the least absolute deviations sense via the $l1$-norm.

\section{Results}\label{sec:results}

\begin{figure*}[t]
    \centering
\setlength{\tabcolsep}{1pt}
\begin{tabular}{ccccc}
Kendall (ref.) & Pearson & Pearson (opt.\ CGLS) & Residual & Residual (opt.\ CGLS) \\
\includegraphics[width=0.195\linewidth,valign=m]{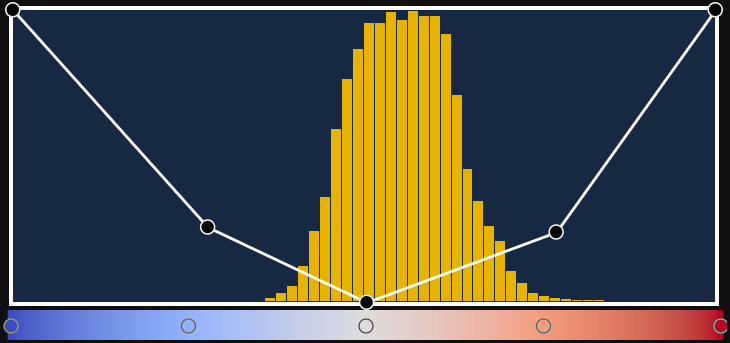} & 
\includegraphics[width=0.195\linewidth,valign=m]{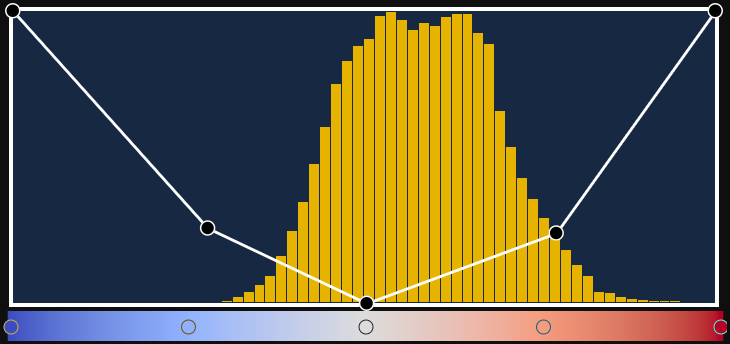} & \includegraphics[width=0.195\linewidth,valign=m]{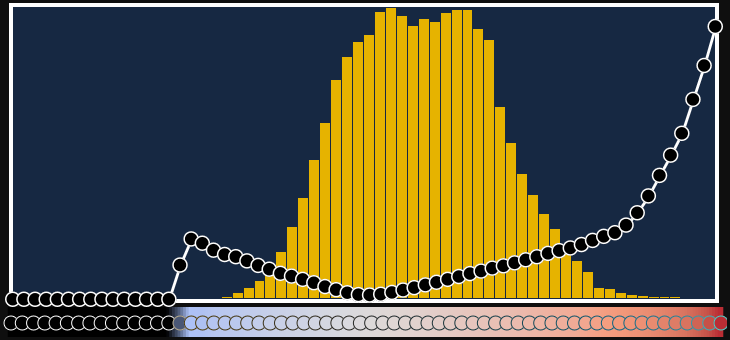} &
\includegraphics[width=0.195\linewidth,valign=m]{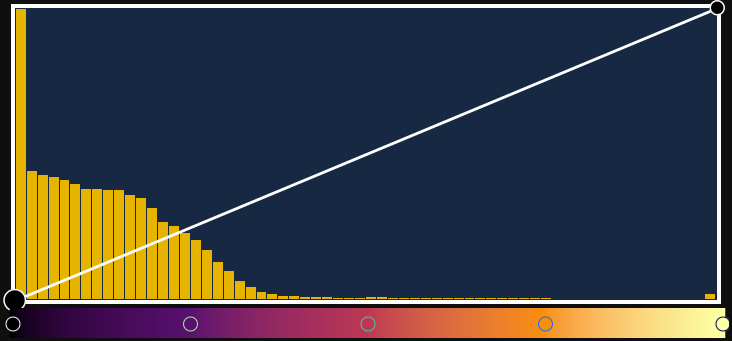} & 
\includegraphics[width=0.195\linewidth,valign=m]{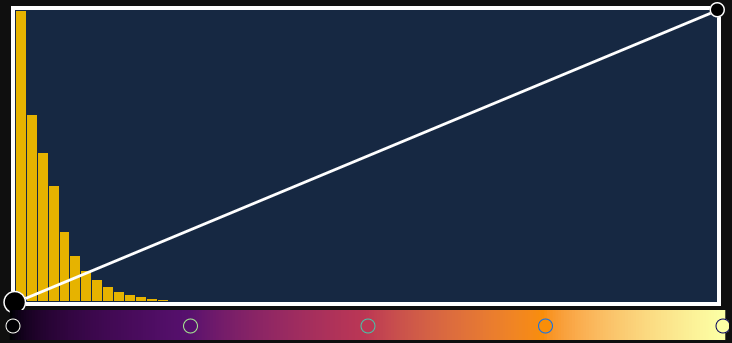} \vspace{0.1cm} \\
\includegraphics[width=0.195\linewidth,valign=m]{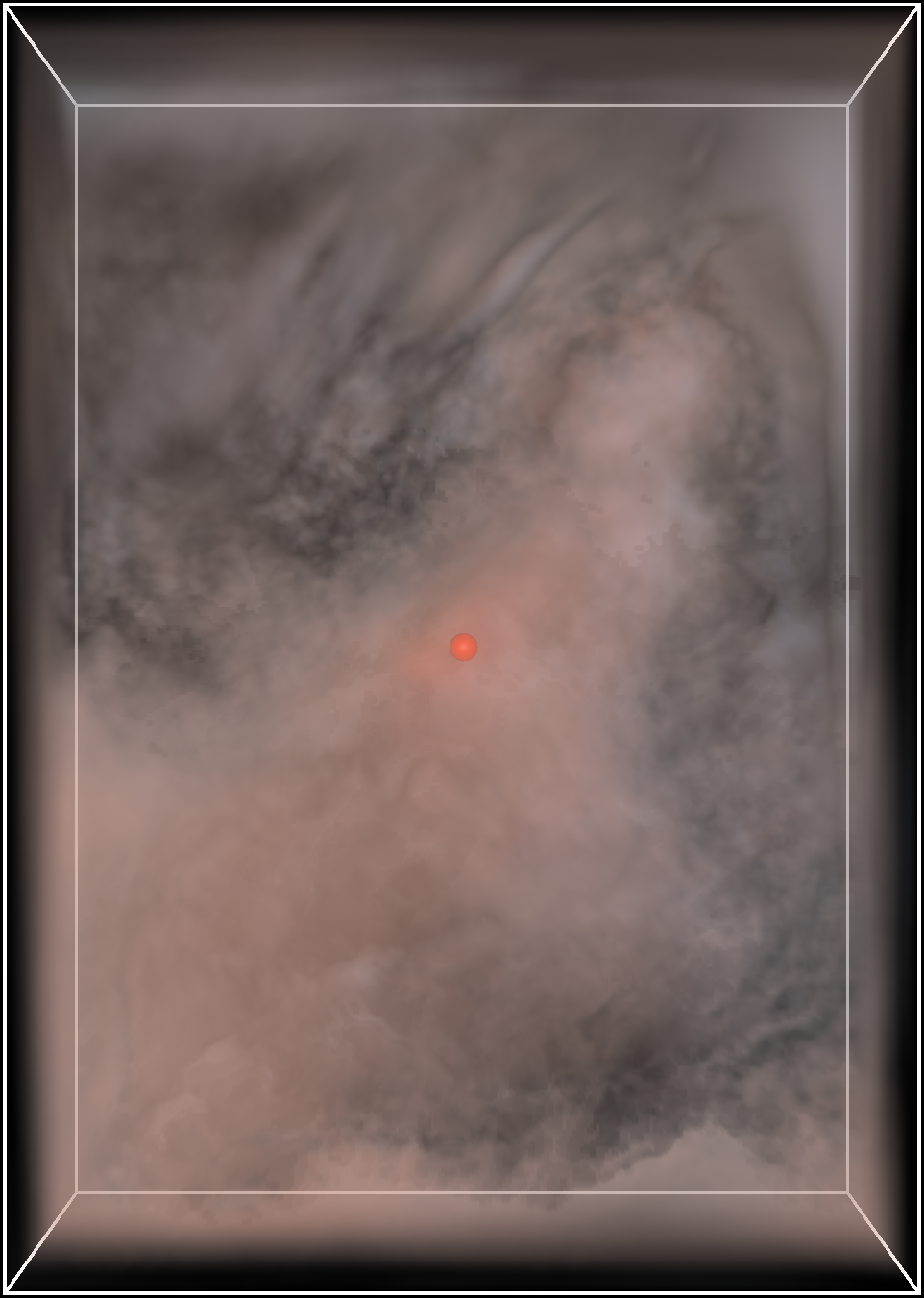} & 
\includegraphics[width=0.195\linewidth,valign=m]{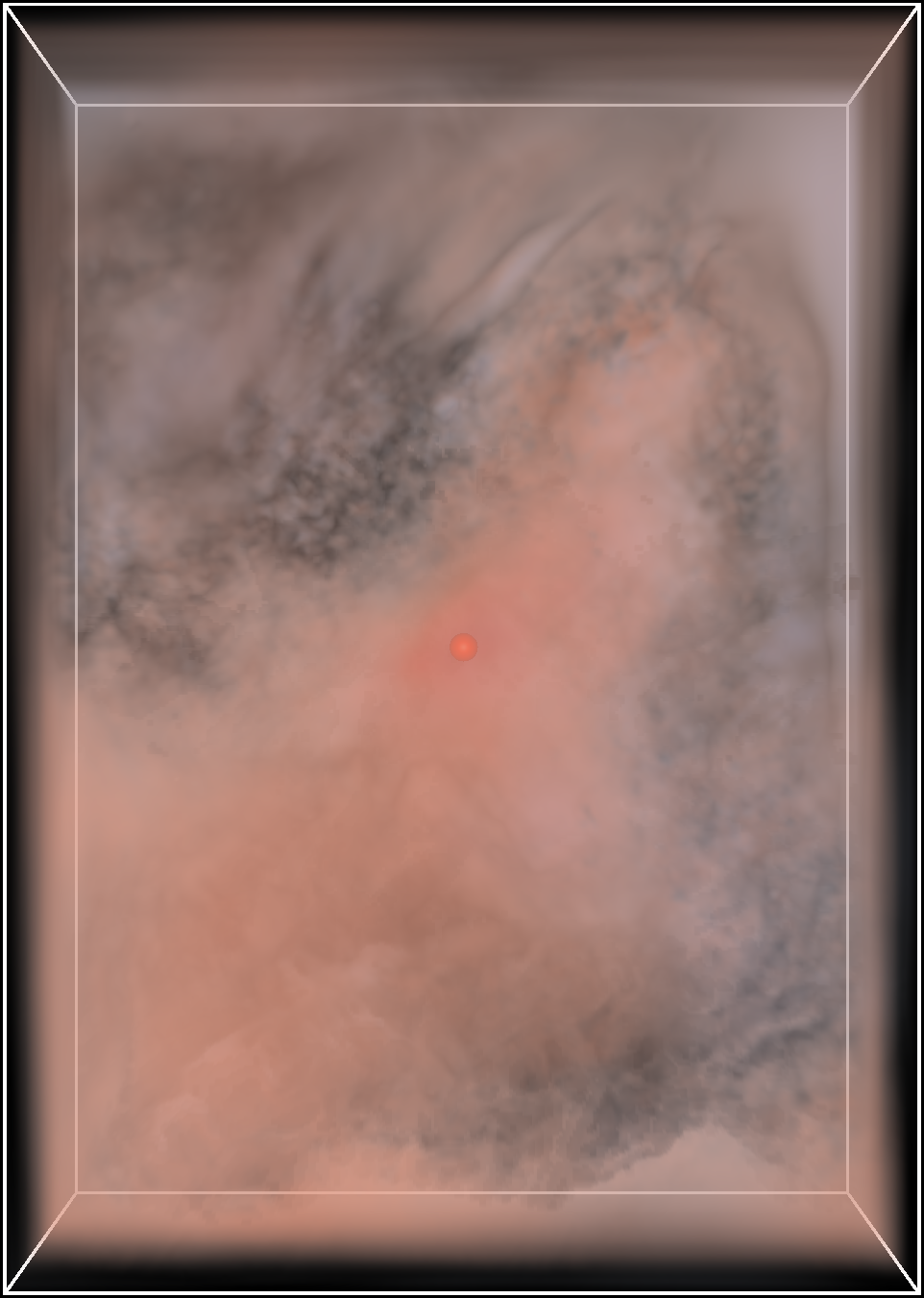} & \includegraphics[width=0.195\linewidth,valign=m]{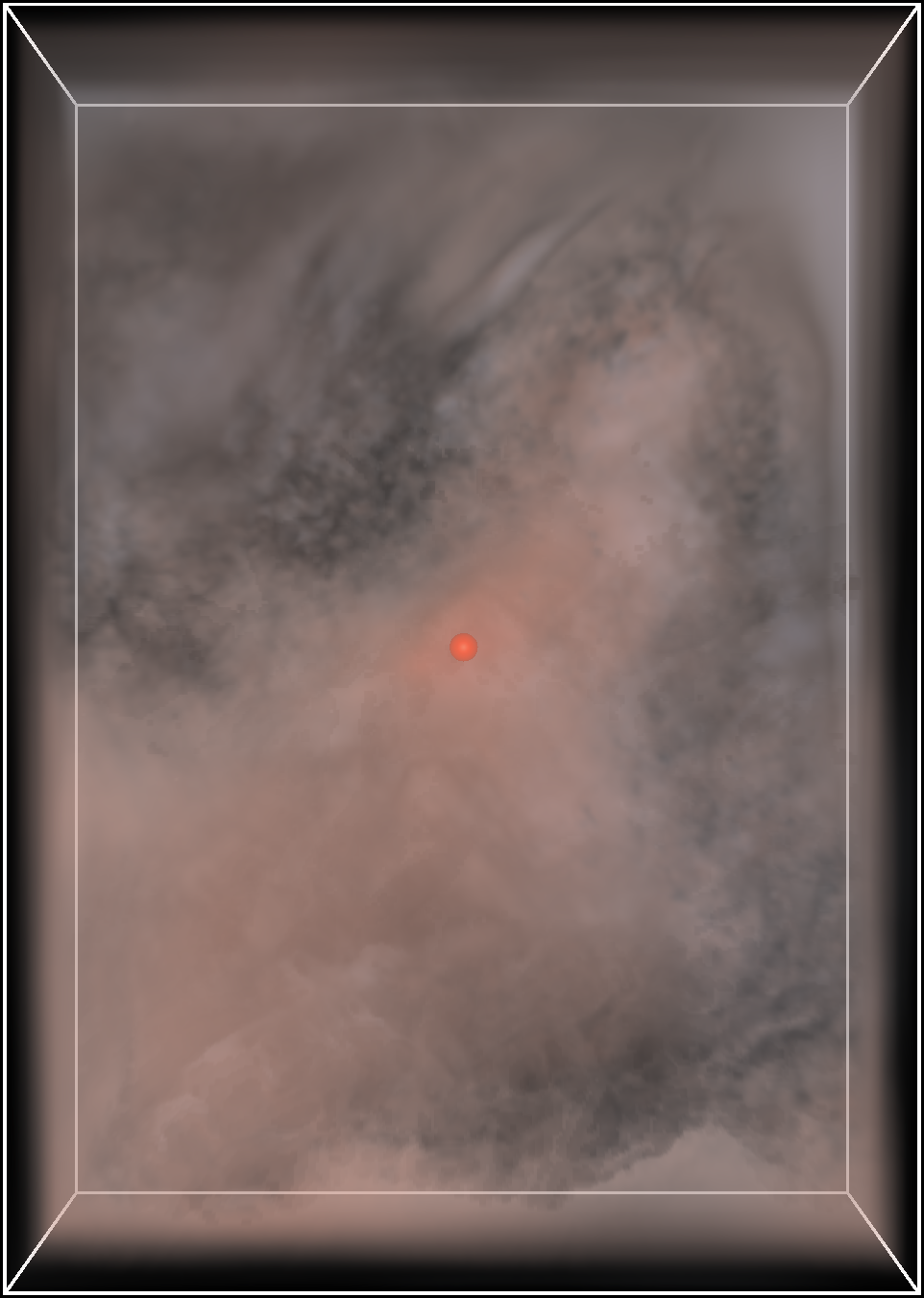} & 
\includegraphics[width=0.195\linewidth,valign=m]{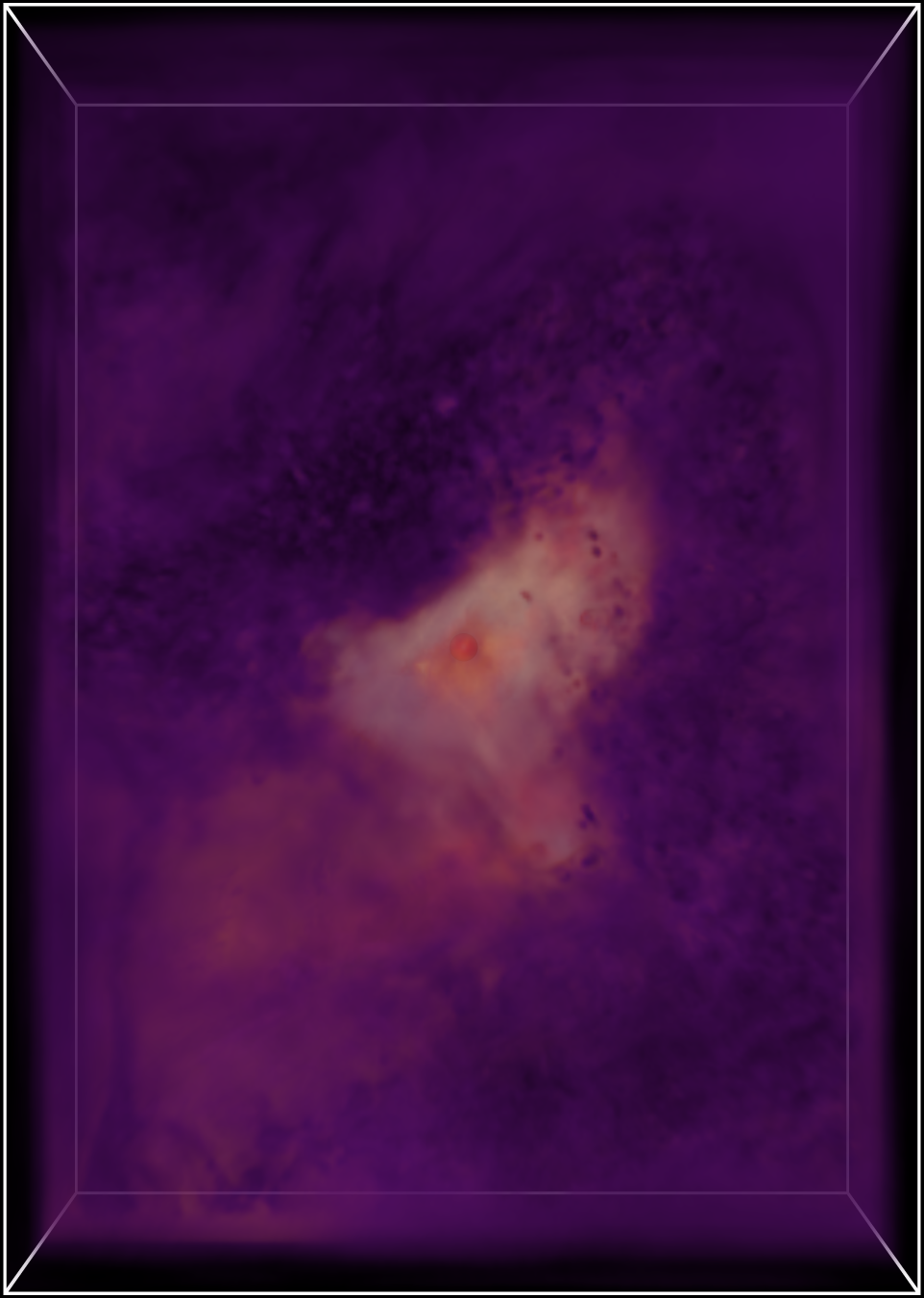} & \includegraphics[width=0.195\linewidth,valign=m]{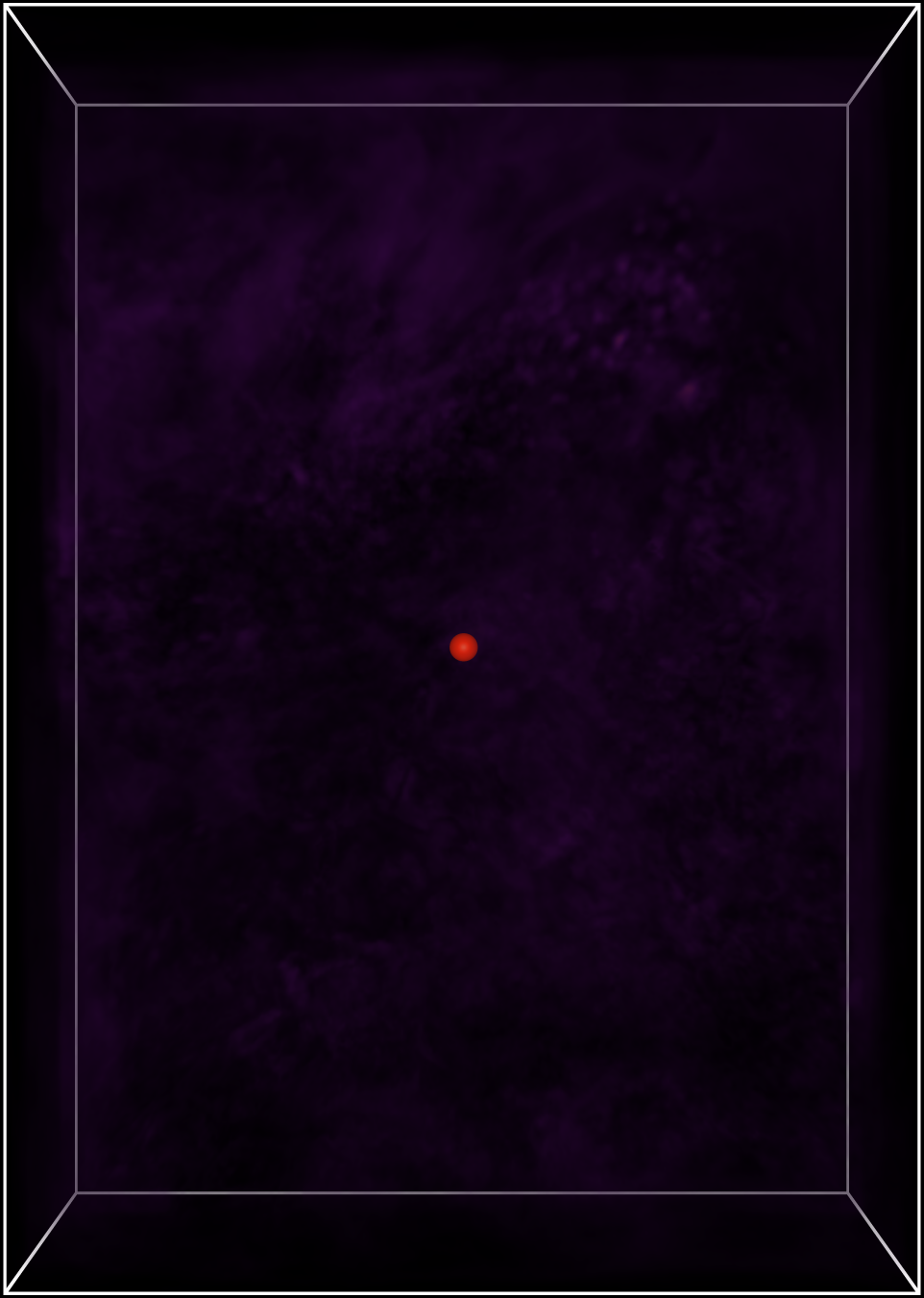} \\
\end{tabular}
    \caption{Transfer function optimization for the comparison of Kendall rank and Pearson correlation coefficients. Top: Reference and optimized transfer functions. Linear transfer functions are used for the residual maps. Bottom: Fields rendered with the corresponding transfer functions with and without optimization. The red dot shows the reference point to which all correlations are computed. Residual maps indicate the spatial differences between the two data sets.  
    }
    \label{fig:necker-pearson-kendall}
\end{figure*}

In the following, we shed light on the performance and quality of different solvers for computing a transfer function using voxel- and image-based optimization. We perform the analysis with three different datasets.
\begin{itemize}
\item Necker et al.~\cite{Necker2020}: A large forecast ensemble comprising 1000 ensemble members of a convective-scale weather forecast over central Europe. The values are stored on a regular grid of size $250 \times 352 \times 20$.
\item Combustion: A turbulent combustion data set on a regular grid of size $480 \times 720 \times 120$. We use the two variables heat release (\texttt{hr}) and mass fraction of the hydroxyl radical (\texttt{Y\_OH}) for comparative visualization.
\item Matsunobu et al.~\cite{MatsunobuEnsemble}: An instance of a weather ensemble, which has been generated using the ICON-D2 numerical weather prediction model. We select the longitudinal and latitudinal wind components \texttt{u} and \texttt{v} on the 11th of August 2020 at 00:00 for ensemble member $1$. The data is stored on a grid of size $651 \times 716 \times 65$.  
\end{itemize}

\subsection{Comparative Visualization Scenarios}\label{sec:demonstration}

In the following, we demonstrate the results and limits of transfer function optimization for the comparative visual analysis of two scalar fields. We show the initial transfer functions $T_r$ and the optimized transfer functions $T_o$ on top of the corresponding images. Histograms of the scalar fields are shown in yellow, and the opacity transfer function is drawn on top. Below each histogram, a color bar indicates the color mapping.

In the first example, we analyze two different measures for indicating correlations in the data set by Necker et al.\ (see \autoref{fig:necker-pearson-kendall}). 
For a selected point in the domain (colored red in the figures), spatial correlations for the temperature \texttt{tk} to all other points in the domain are computed over the ensemble axis, and the resulting spatial correlation fields are rendered.  
We compute two different measures of spatial dependence, the Pearson product-moment correlation coefficient (PPMCC)~\cite{Pearson1,Pearson2} and the Kendall rank correlation coefficient (KRCC)~\cite{Kendall}.
The PPMCC measures the strength of linear correlation, and the KRCC measures the ordinal association between two sets of data. 
To shed light on the structural differences between the two correlation measures, they are first rendered using the same transfer function $T_r$ (1st and 2nd columns in \autoref{fig:necker-pearson-kendall}). As can be seen, the resulting images look significantly different.   
When using the optimized transfer function (3rd column in \autoref{fig:necker-pearson-kendall}) generated with any least squares solver, the image of PPMCC now looks very similar to the image of PPMCC. This is further confirmed by the residual map between two measures using the reference and optimized transfer function (4th and 5th column in \autoref{fig:necker-pearson-kendall}). Thus, it can be concluded that both PPMCC and KRCC, for the selected point, are similar concerning the distribution of values across the domain. Note here that an interactive visual inspection of the residual map enables a far more fine granular analysis of the spatial differences.


\begin{figure*}[!h]
    \centering
\setlength{\tabcolsep}{1pt}
\begin{tabular}{ccccc}
Y\_OH (ref.) & hr & hr (opt. CGLS) & hr (opt. DiffDVR $l2$) & hr (opt. DiffDVR $l1$) \\
\includegraphics[width=0.195\linewidth,valign=m]{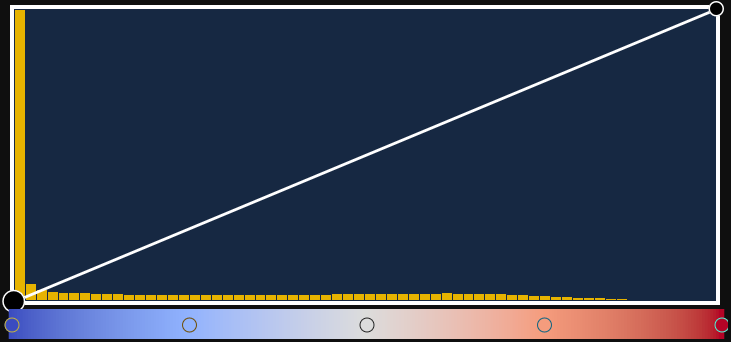} & \includegraphics[width=0.195\linewidth,valign=m]{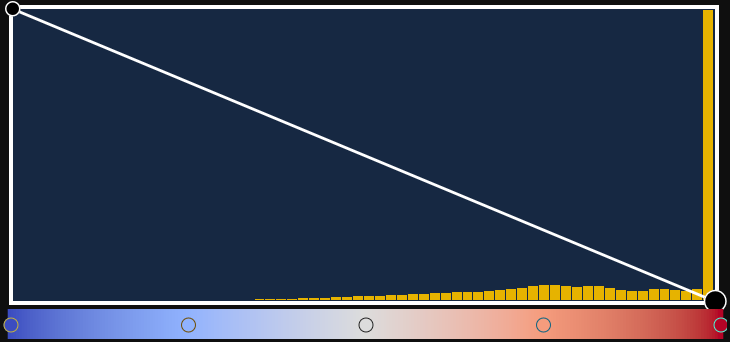} & \includegraphics[width=0.195\linewidth,valign=m]{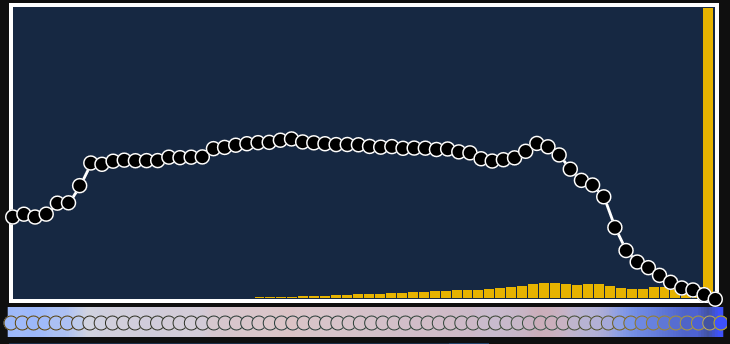} & \includegraphics[width=0.195\linewidth,valign=m]{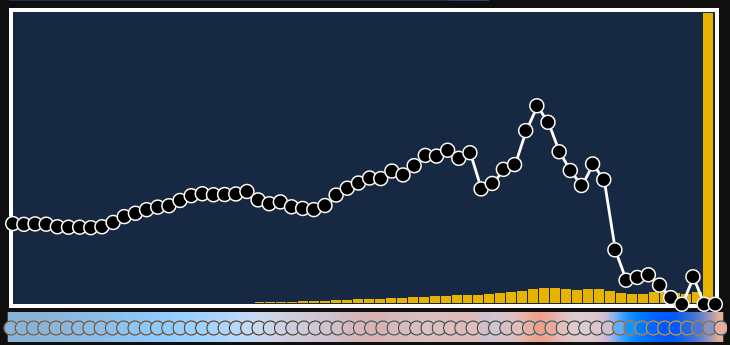} &
\includegraphics[width=0.195\linewidth,valign=m]{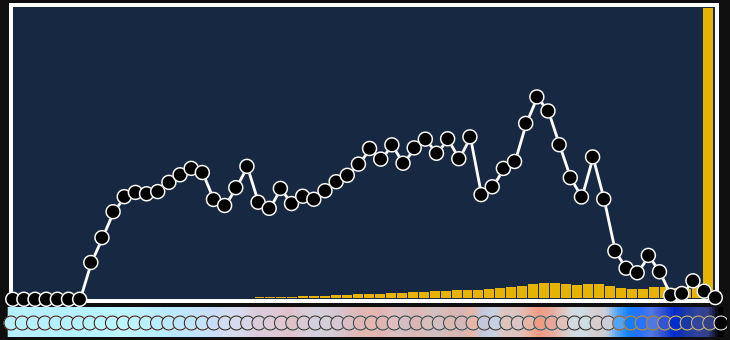} \vspace{0.1cm} \\
\includegraphics[width=0.195\linewidth,valign=m]{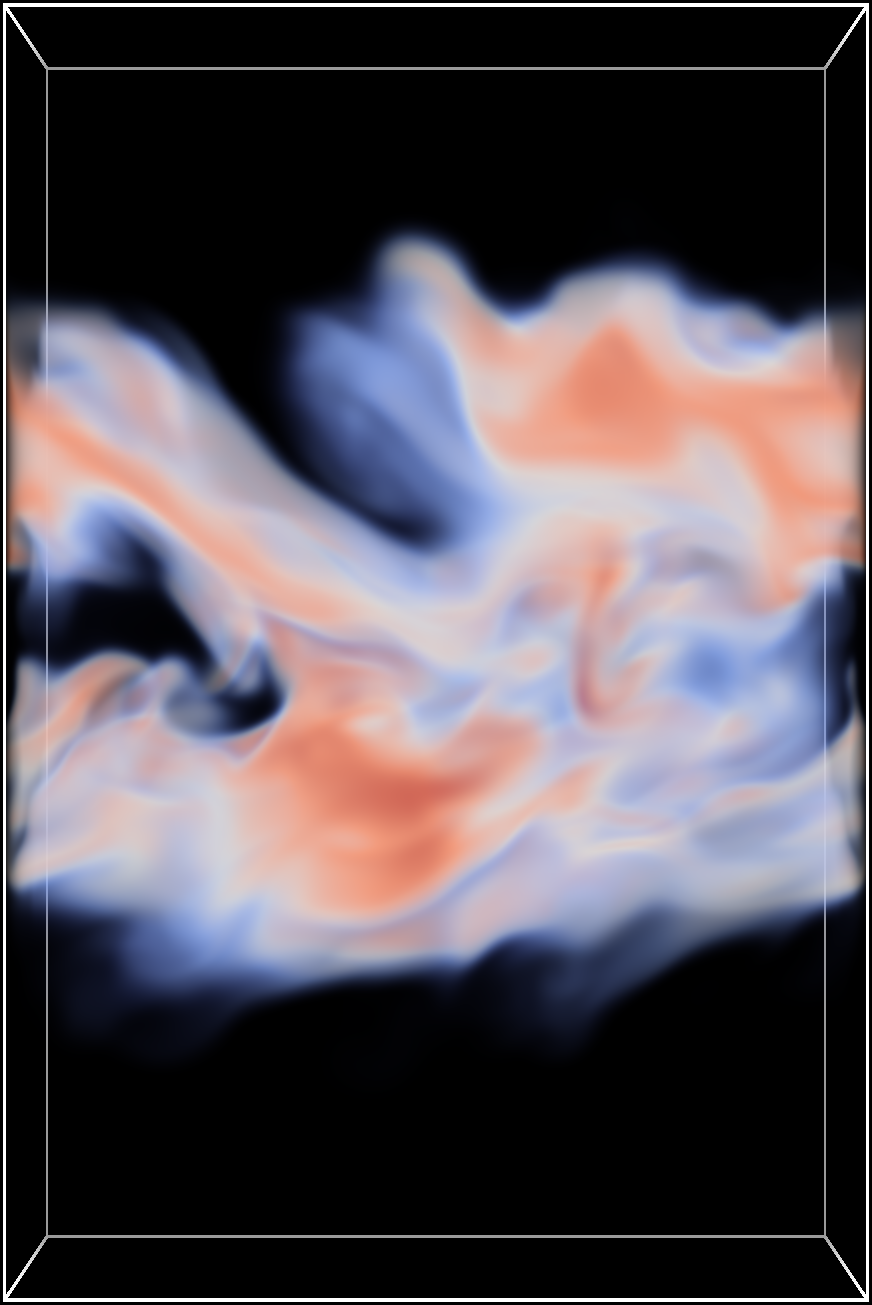} & \includegraphics[width=0.195\linewidth,valign=m]{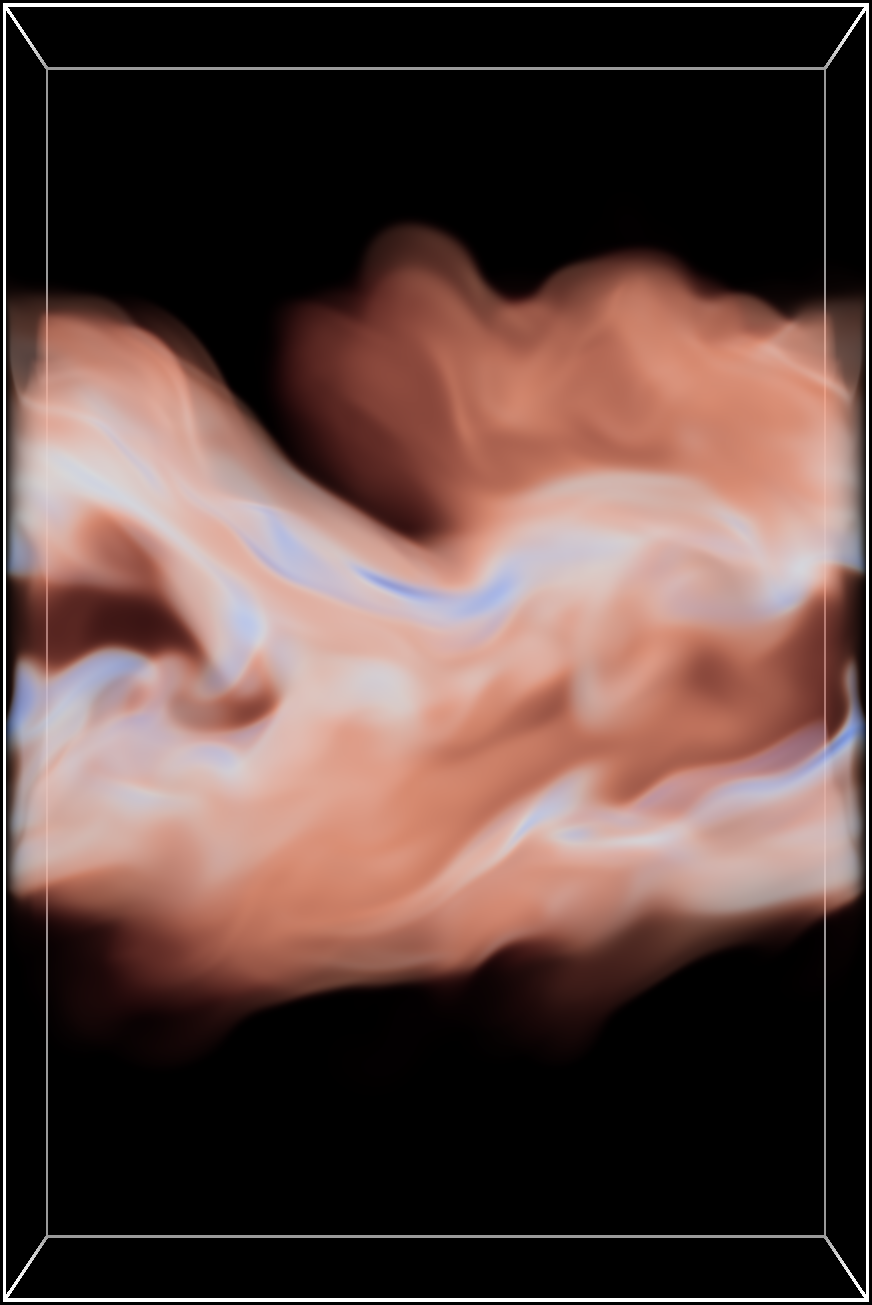} & \includegraphics[width=0.195\linewidth,valign=m]{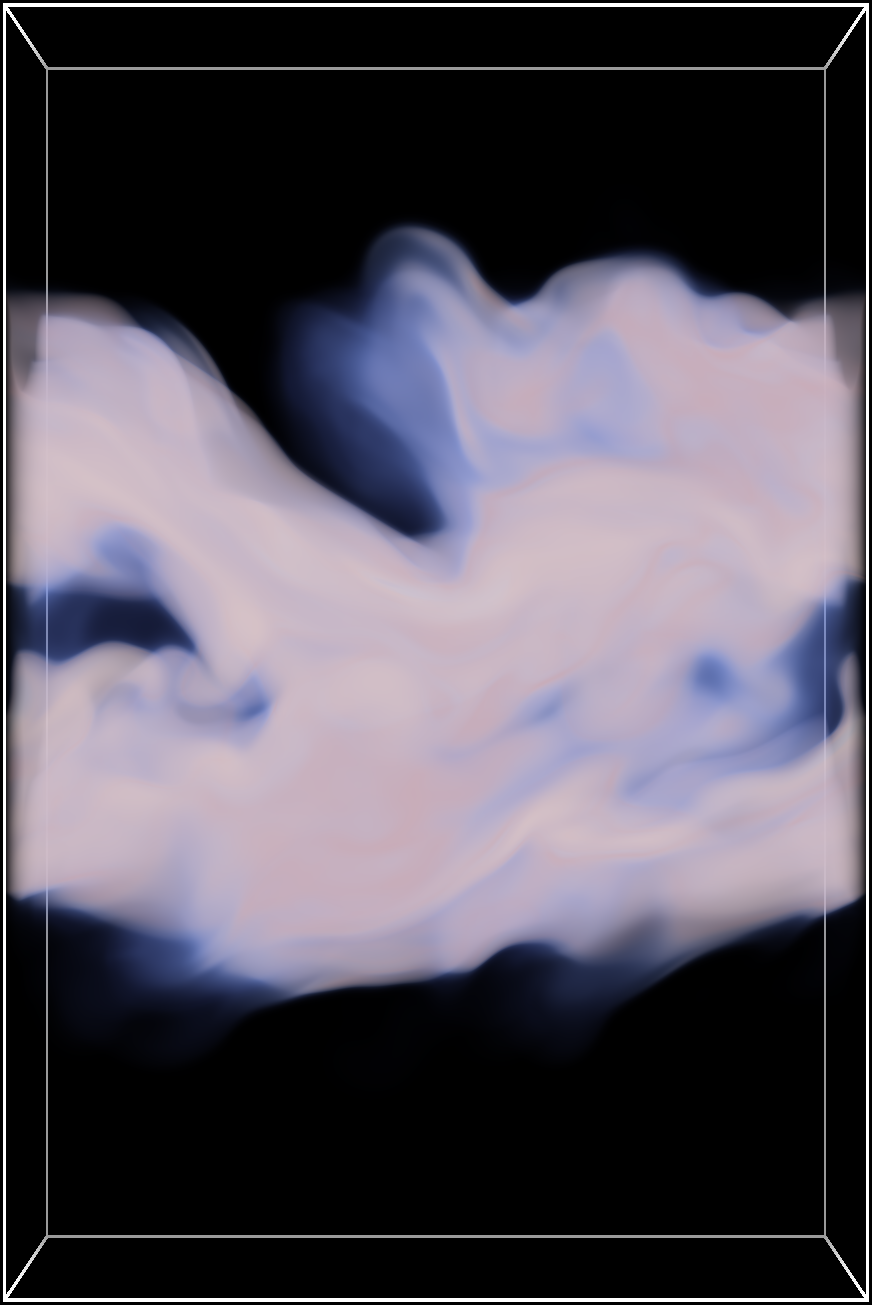} & \includegraphics[width=0.195\linewidth,valign=m]{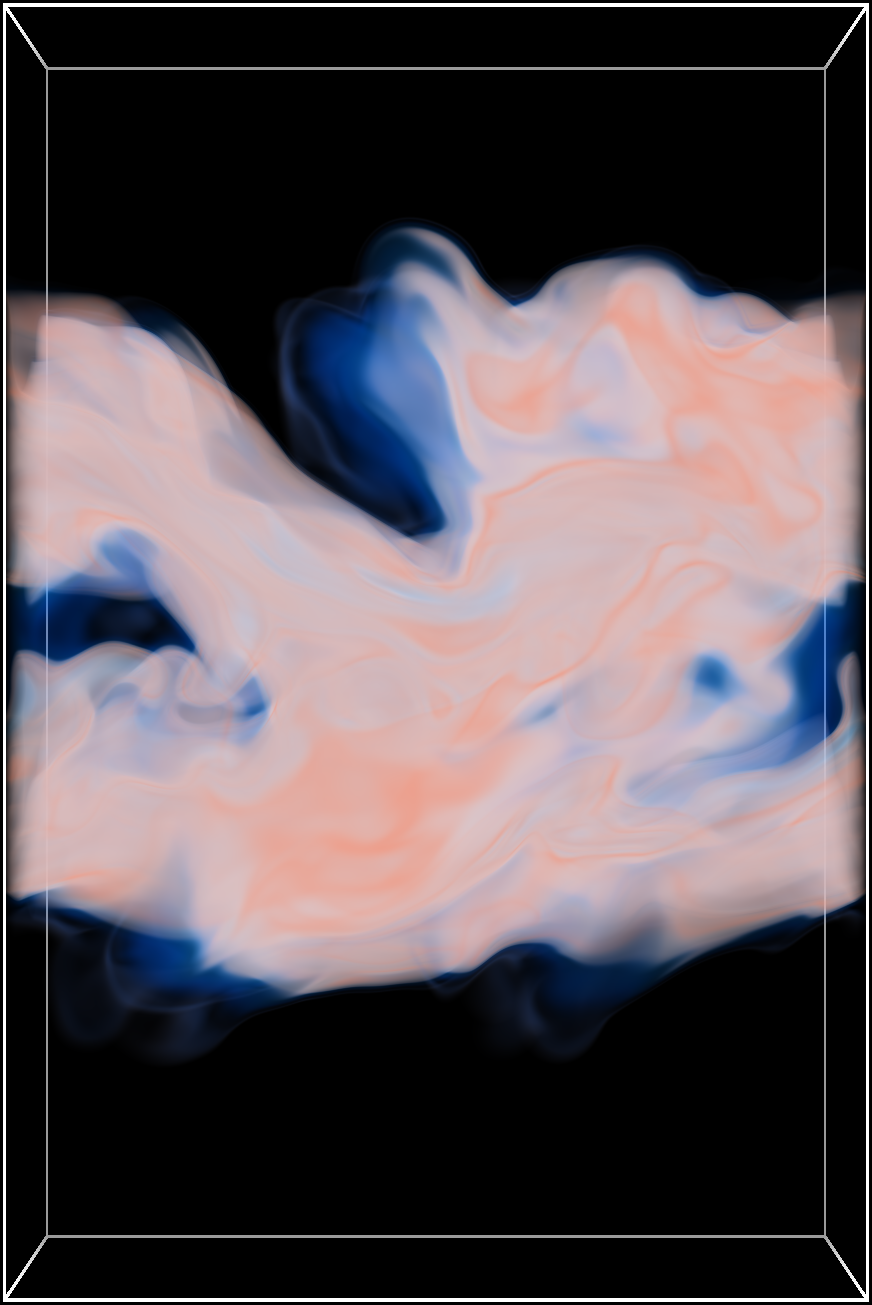} &
\includegraphics[width=0.195\linewidth,valign=m]{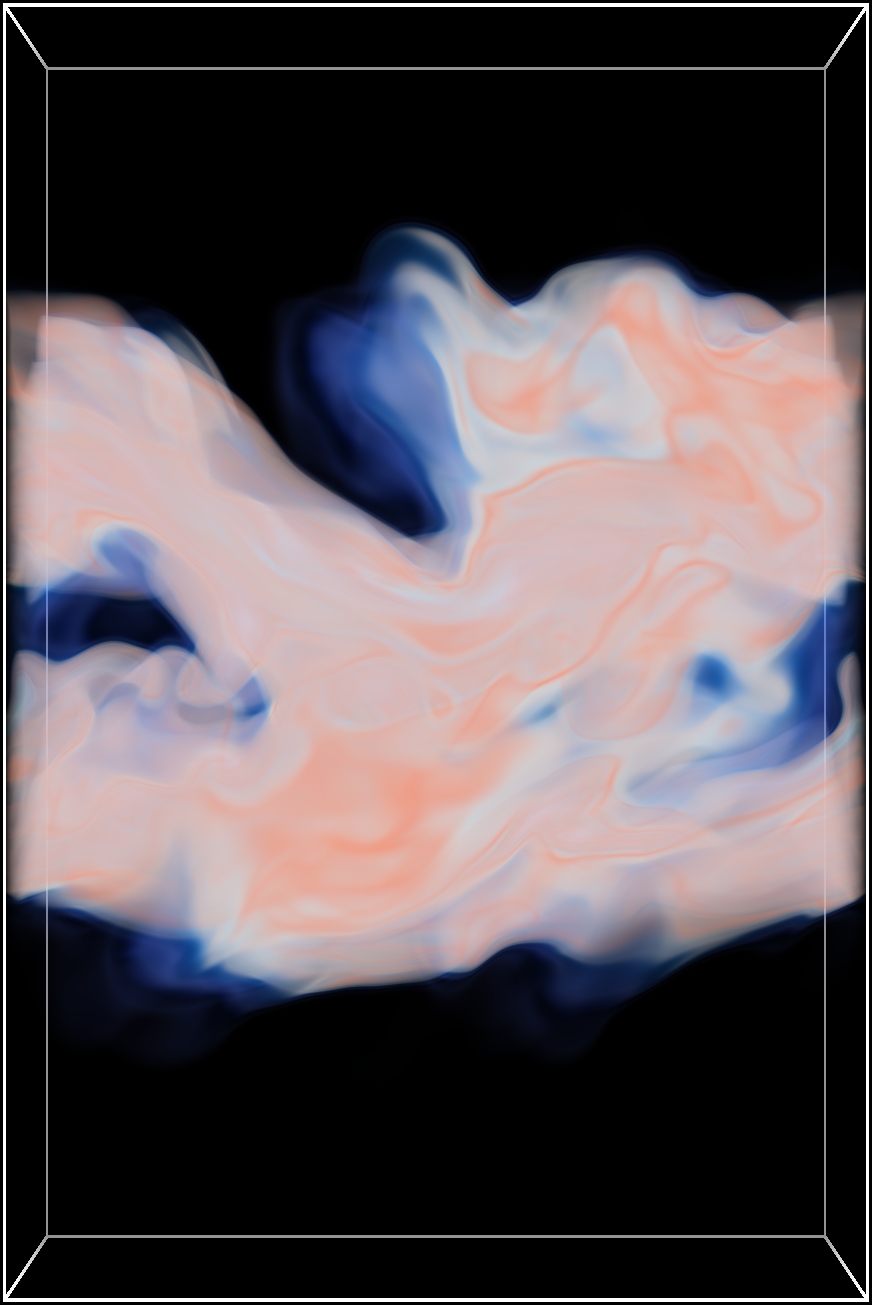} \vspace{0.1cm} \\
Voxel color residuals:& \includegraphics[width=0.195\linewidth,valign=m]{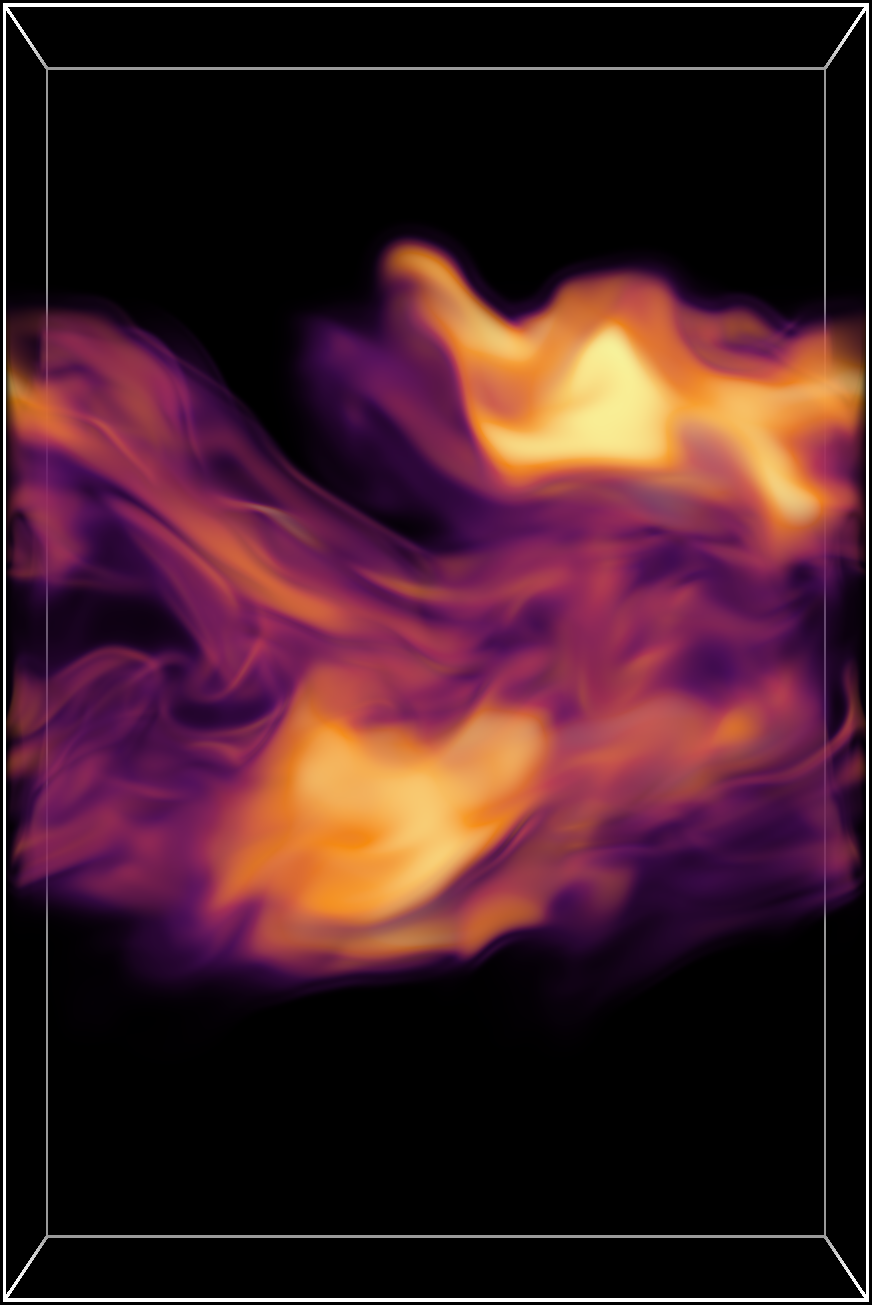} & \includegraphics[width=0.195\linewidth,valign=m]{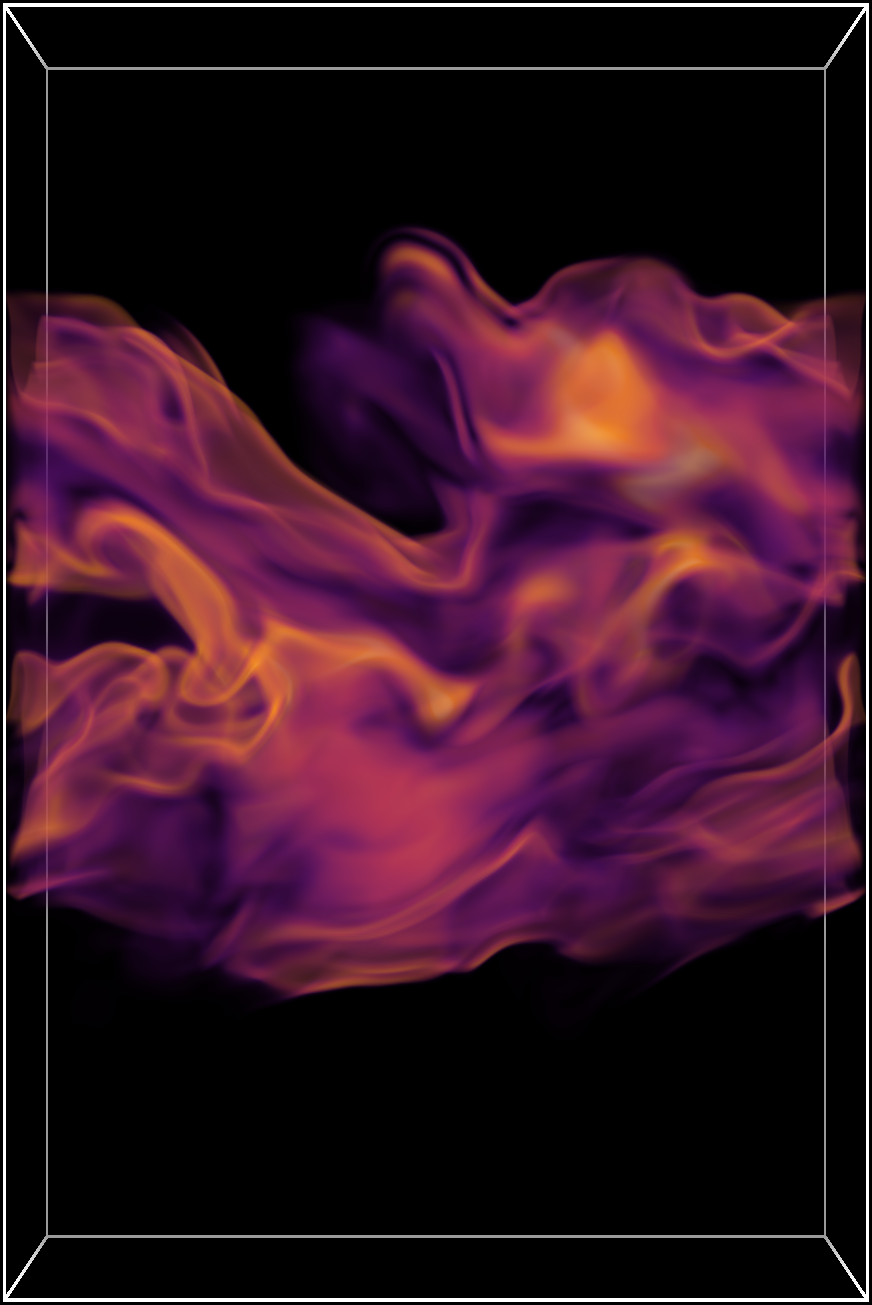} & \includegraphics[width=0.195\linewidth,valign=m]{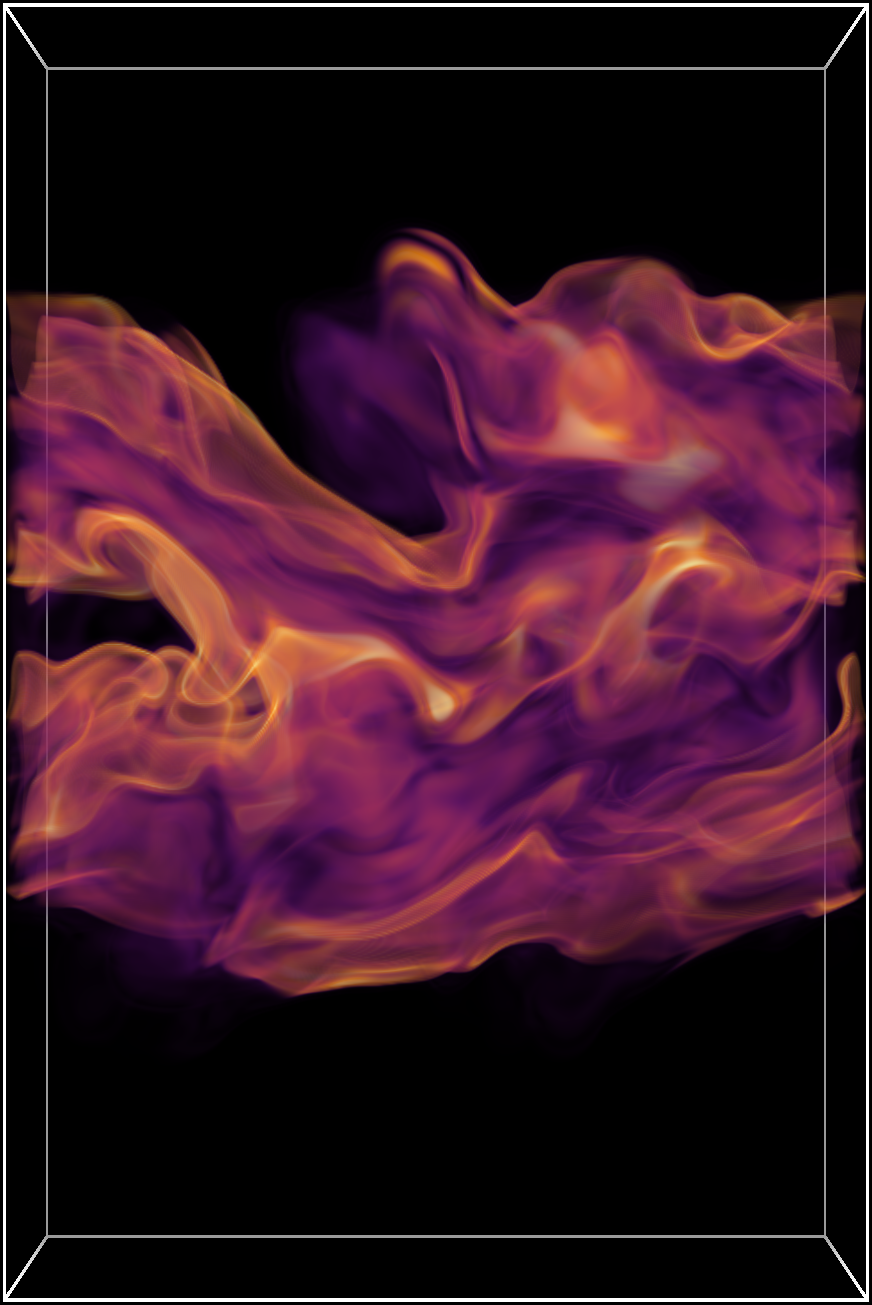} &
\includegraphics[width=0.195\linewidth,valign=m]{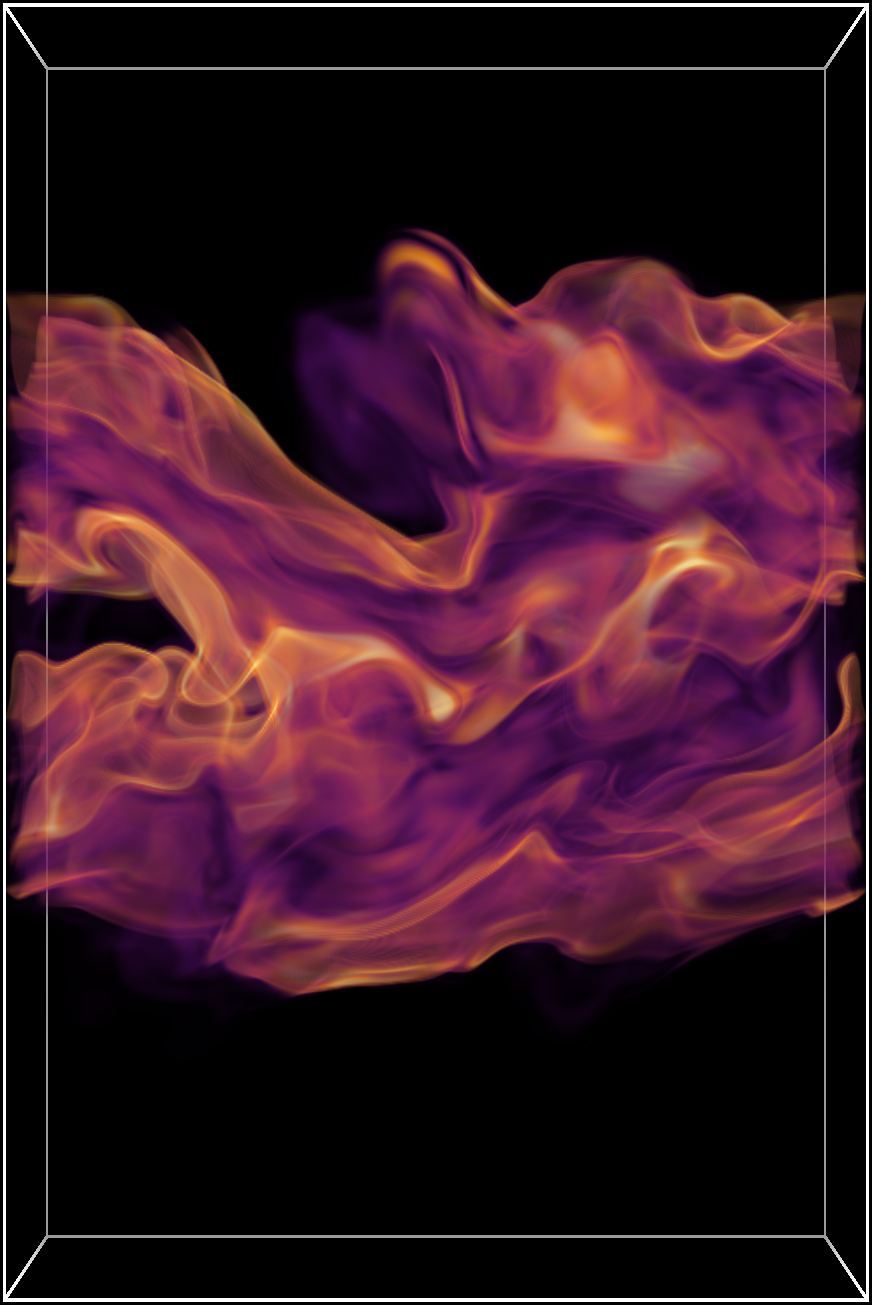}\\
\end{tabular}
    \caption{Transfer function optimization for variables \texttt{Y\_OH} and \texttt{hr} in a turbulent combustion simulation. Corresponding residual maps to the reference field \texttt{Y\_OH} are shown at the bottom.}
    \label{fig:combustion}
\end{figure*}

In \autoref{fig:combustion}, the two selected variables from the turbulent combustion simulation are compared. The transfer function optimized via the ordinary least squares (OLS) approach and the corresponding image of the variable \texttt{hr}  are shown in the 3rd column. Notably all solvers operating on voxel colors (normal equations-based, quadratic programming-based, etc.) converge towards the same optimization result. The results of optimization using DiffDVR, using $L_2$ and $L_1$ loss, seem to match the target image of variable \texttt{Y\_OH} more closely. This is also confirmed by the residual maps in the last row, where DiffDVR seems to generate a transfer function that lets the rendered images appear less washed out. 
The image-based loss more aggressively penalizes differences in voxels that are more visible to the viewer. 
Interestingly, while the residual also decreases after optimization, it does not converge to low values across the domain. This hints on some global structural dissimilarity between the two scalar field distributions which cannot be compensated by the optimized transfer function. 

In \autoref{tab:combustion-img-metrics}, we further analyse the similarity of the images of the two simulation parameters in \autoref{fig:combustion}. The visual similarity does not reflect in these metrics, as it seems the more smoothed out the results are (when using the CGLS solver) the higher their similarity. The more crisp results when applying the transfer function optimized via DiffDVR are deemed least similar. The choice of the loss function ($l1$ vs. $l2$) can make a minor difference in the final results. The $l2$-norm is known to penalize larger residuals more strongly due to the quadratic term, while the $l1$-norm may tolerate some larger outliers in favor of better minimization of many already lower deviations (cf. results by Zhao et al.~\cite{l1loss}). 

\renewcommand{\tabularxcolumn}[1]{m{#1}}
\renewcommand{\arraystretch}{1.4}
\begin{table}[ht]
        \setlength{\tabcolsep}{4pt}
	\fontsize{8pt}{8pt}\selectfont
	\centering%
	\begin{tabularx}{1.0\linewidth}{|X<{\centering}|m{14mm}<{\centering}|m{14mm}<{\centering}|m{14mm}<{\centering}|m{14mm}<{\centering}|}%
		\hline
		\multicolumn{1}{|>{\centering}c|}{\textbf{Solver}} 
		& \multicolumn{1}{>{\centering}c|}{\textbf{Original}} 
		& \multicolumn{1}{>{\centering}c|}{\textbf{CGLS}}
		& \multicolumn{1}{>{\centering}c|}{\textbf{DiffDVR $l2$}}
		& \multicolumn{1}{>{\centering}c|}{\textbf{DiffDVR $l1$}}
		\\ \hline
		
		RMSE            & 30.49 & 19.51 & 20.72 & 20.28 \\ \hline
		PSNR            & 18.44 & 22.32 & 21.80 & 21.98 \\ \hline
		SSIM            & 0.907 & 0.938 & 0.894 & 0.911 \\ \hline
		LPIPS (AlexNet) & 0.216 & 0.191 & 0.233 & 0.243 \\ \hline
		LPIPS (VGGNet)  & 0.258 & 0.218 & 0.259 & 0.267 \\ \hline

	\end{tabularx}%
	\vspace{0.2cm}
	\caption{Similarity metrics for the images in \autoref{fig:combustion}. For PSNR and SSIM, higher is better. For RMSE and LPIPS, lower is better.}
	\label{tab:combustion-img-metrics}
\end{table}

\begin{figure*}[t]
    \centering
\setlength{\tabcolsep}{1pt}
\begin{tabular}{ccccc}
 Mutual information (ref.) & Pearson & Pearson (opt.\ CGLS) & Residual & Residual (opt.\ CGLS) \\
\includegraphics[width=0.195\linewidth,valign=m]{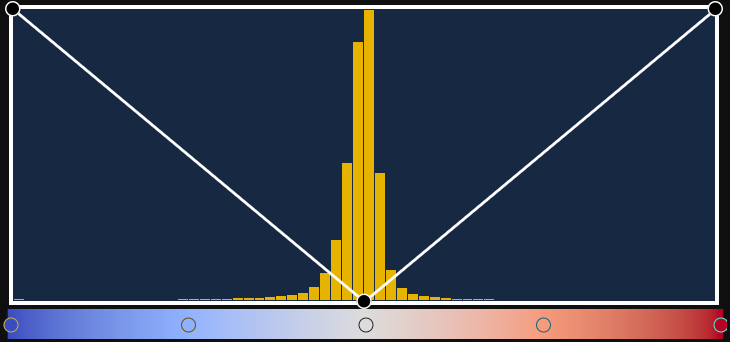} & 
\includegraphics[width=0.195\linewidth,valign=m]{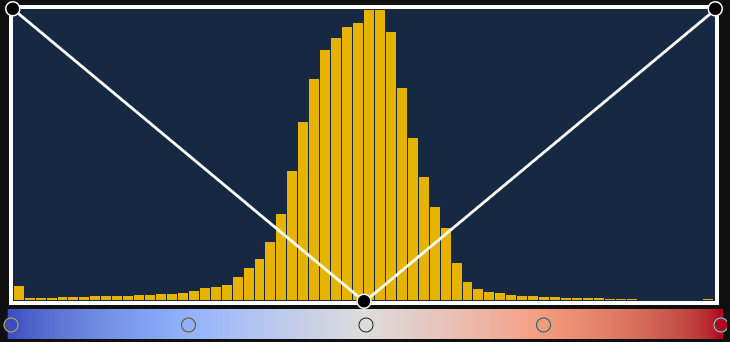} & \includegraphics[width=0.195\linewidth,valign=m]{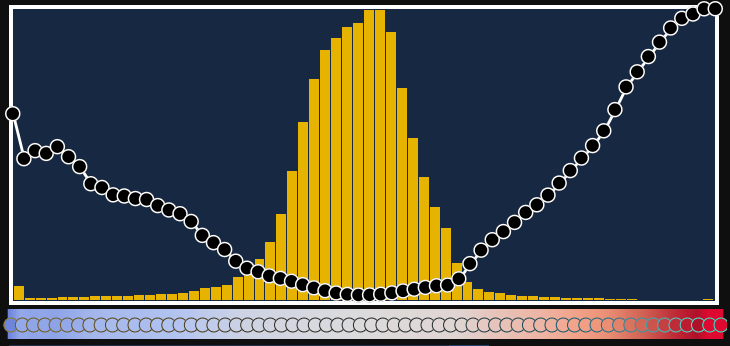} &
\includegraphics[width=0.195\linewidth,valign=m]{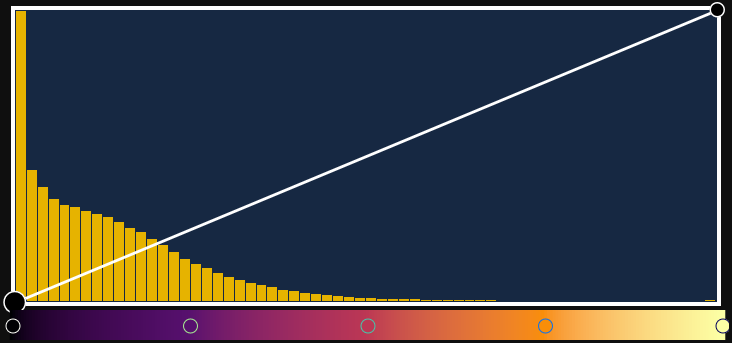} & 
\includegraphics[width=0.195\linewidth,valign=m]{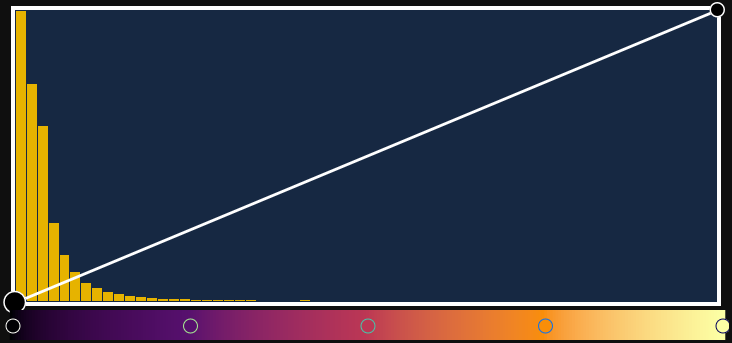} \vspace{0.1cm} \\
\includegraphics[width=0.195\linewidth,valign=m]{figures/Necker/u/mi.png} & 
\includegraphics[width=0.195\linewidth,valign=m]{figures/Necker/u/pearson_orig.png} & \includegraphics[width=0.195\linewidth,valign=m]{figures/Necker/u/pearson_opt.png} & 
\includegraphics[width=0.195\linewidth,valign=m]{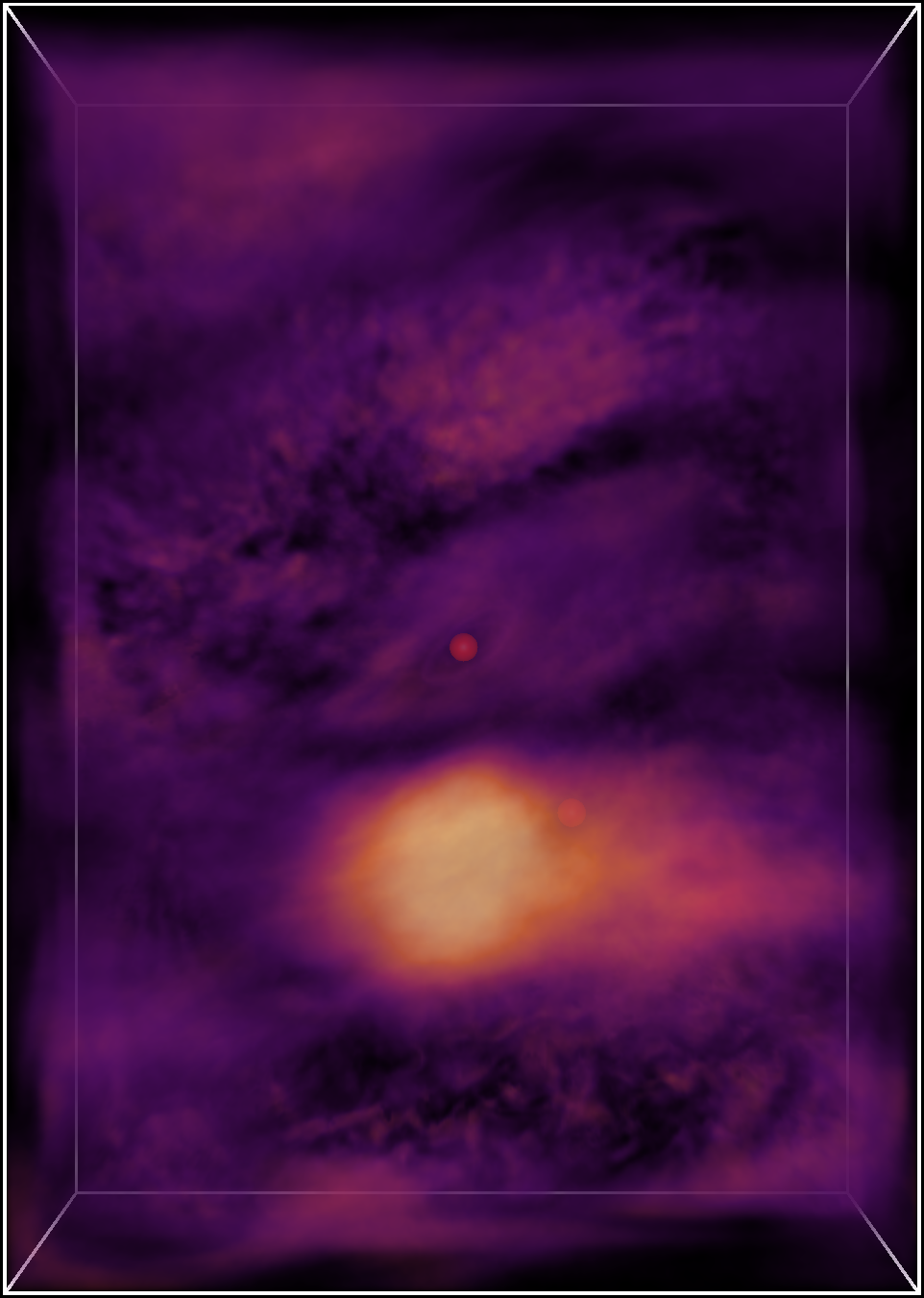} & \includegraphics[width=0.195\linewidth,valign=m]{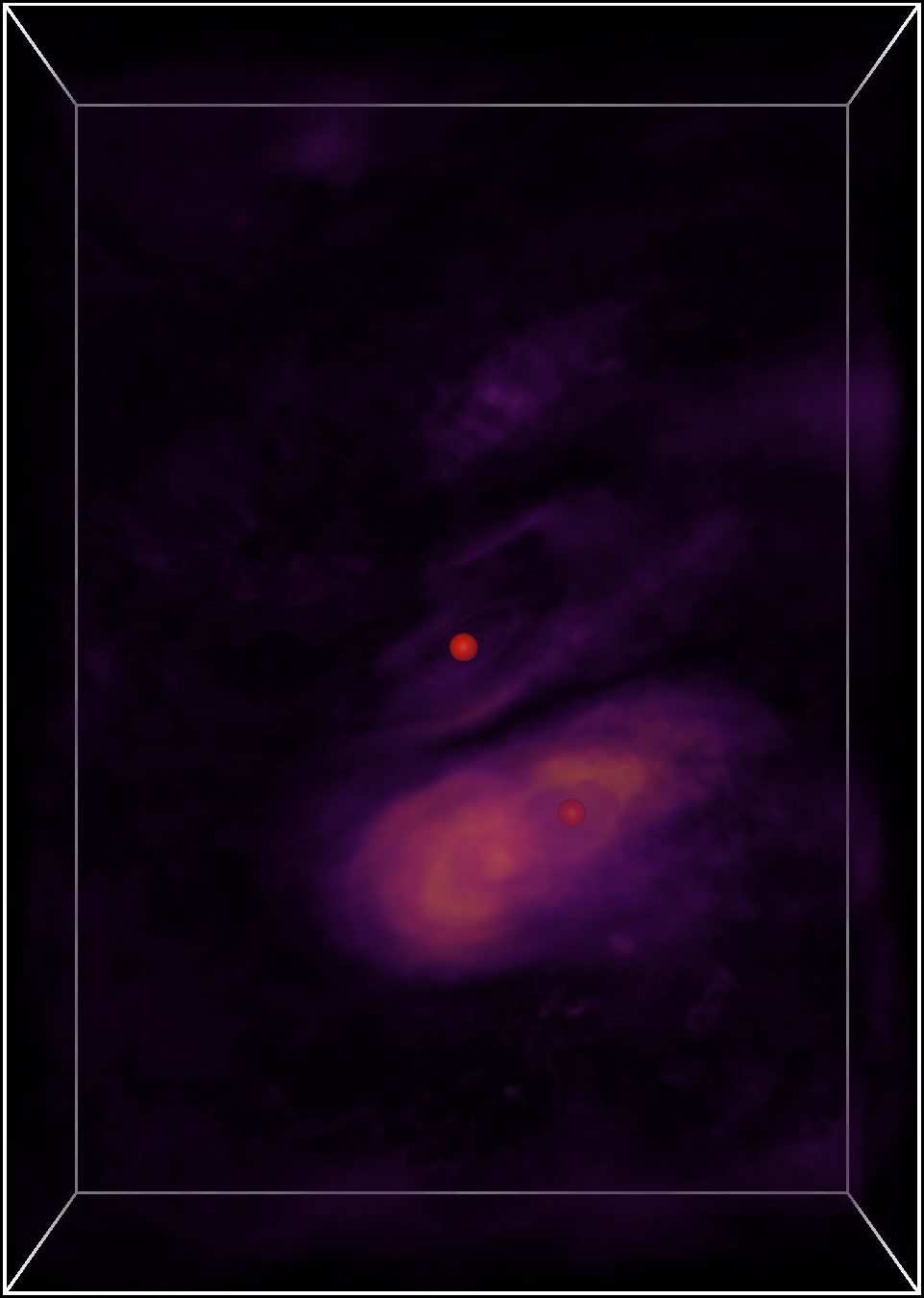} \\
\end{tabular}
    \caption{Same as in \autoref{fig:necker-pearson-kendall}, but now the correlation fields for two points are computed (once using the mutual information measure, and once using the Pearson correlation coefficients), and the difference fields for both measures are compared. 
    The red dots show the reference points for correlation computation.}
    \label{fig:necker-pearson-mi}
    \centering
\setlength{\tabcolsep}{1pt}
\begin{tabular}{ccccc}
u (ref.) & v & v (opt.\ CGLS) & Residual & Residual (opt.\ CGLS) \\
\includegraphics[width=0.195\linewidth,valign=m]{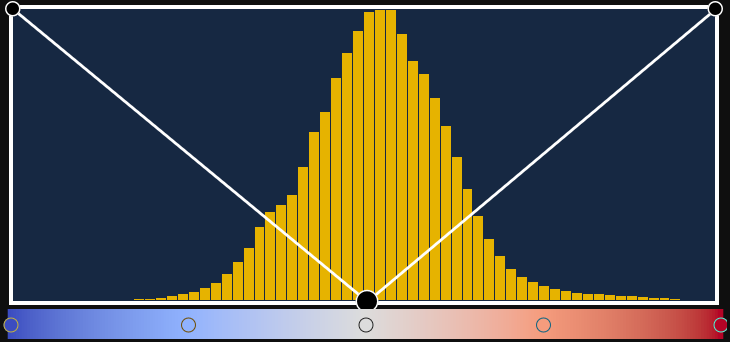} & \includegraphics[width=0.195\linewidth,valign=m]{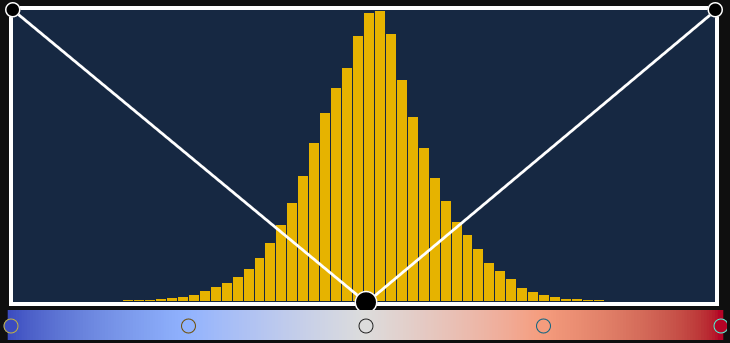} & \includegraphics[width=0.195\linewidth,valign=m]{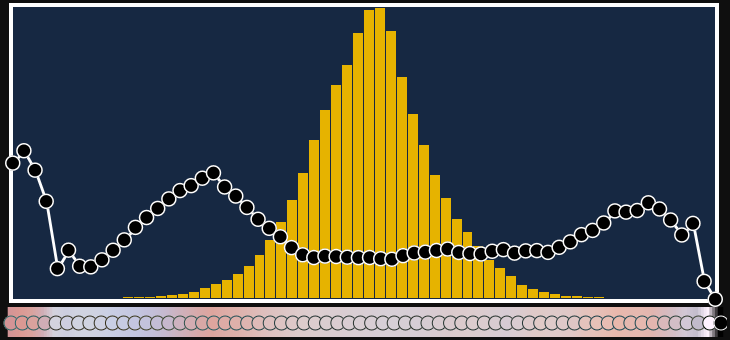} & \includegraphics[width=0.195\linewidth,valign=m]{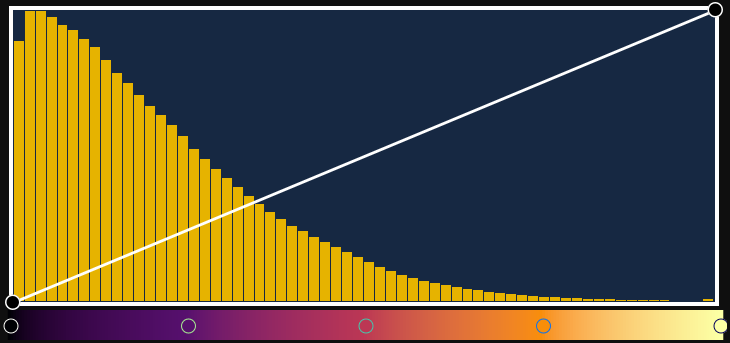} & \includegraphics[width=0.195\linewidth,valign=m]{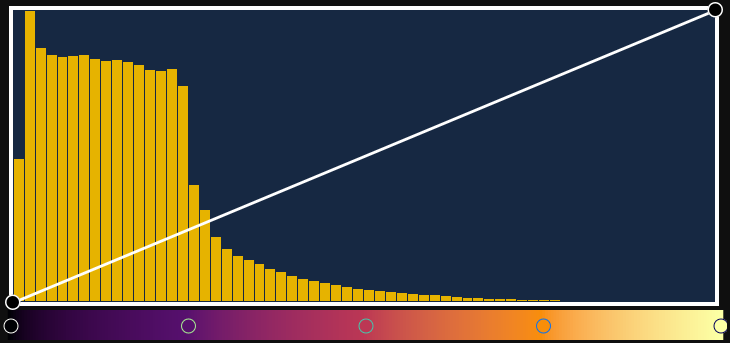} \vspace{0.1cm} \\
\includegraphics[width=0.195\linewidth,valign=m]{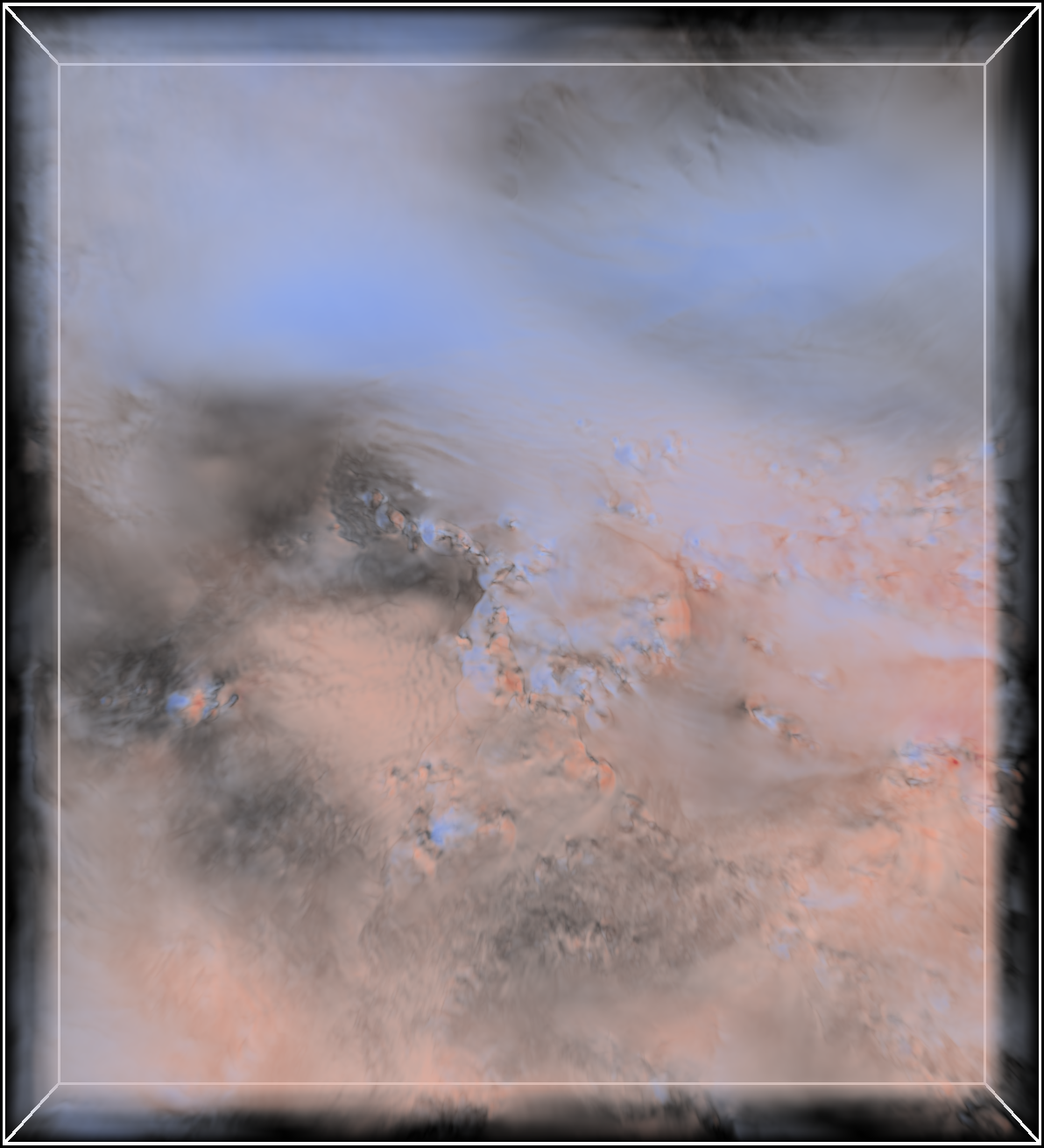} & \includegraphics[width=0.195\linewidth,valign=m]{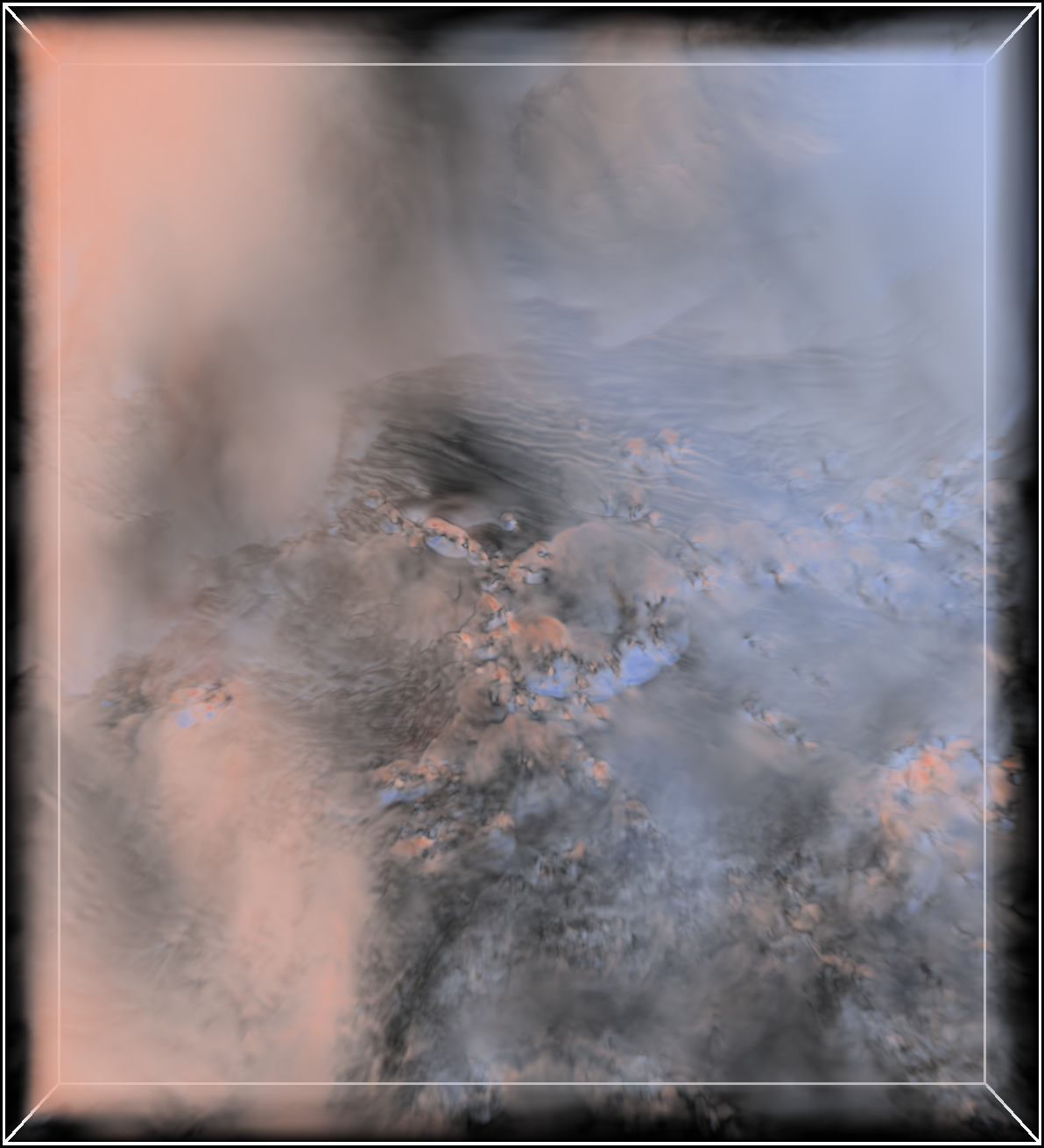} & \includegraphics[width=0.195\linewidth,valign=m]{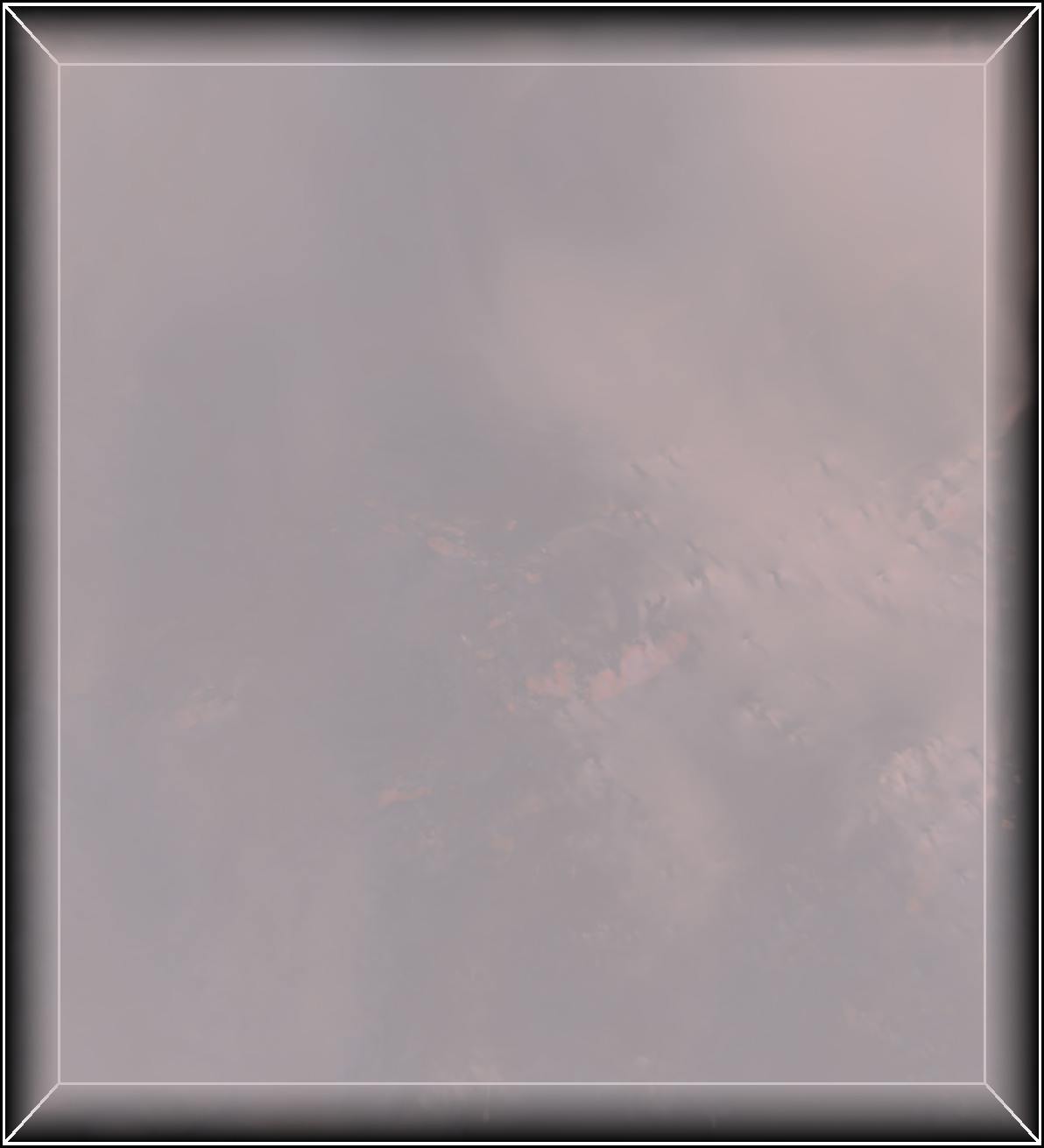} & \includegraphics[width=0.195\linewidth,valign=m]{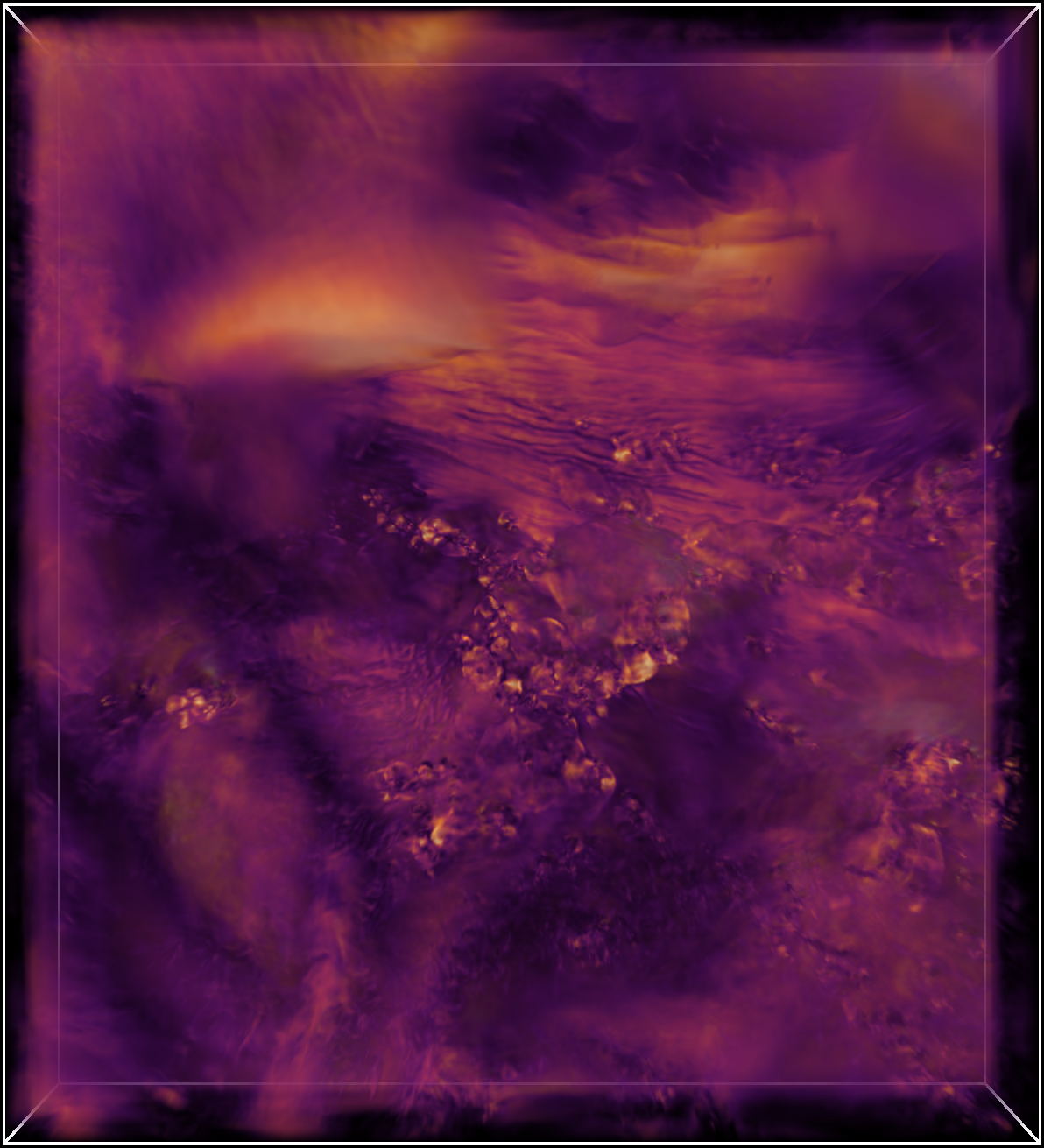} & \includegraphics[width=0.195\linewidth,valign=m]{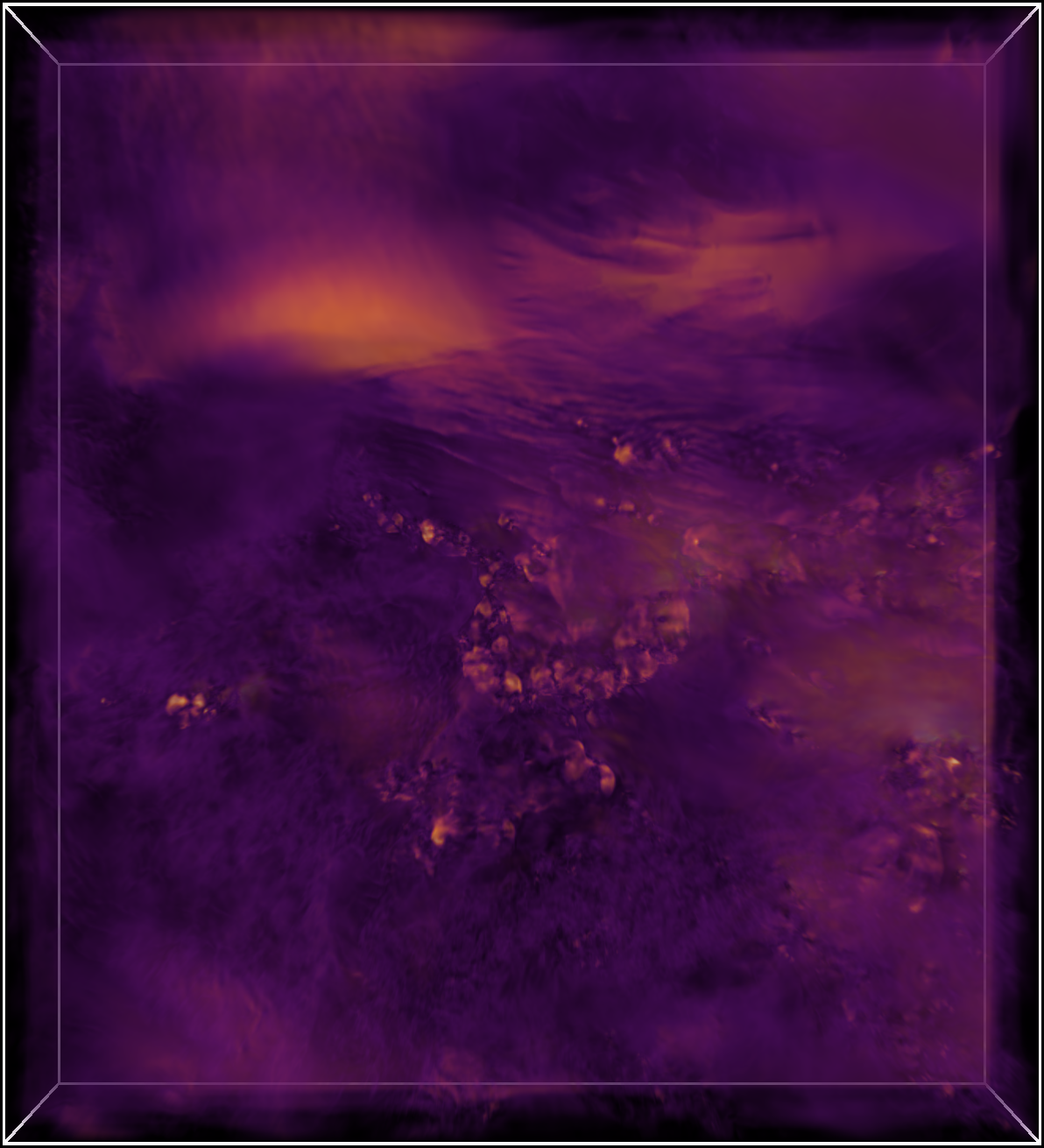} \\
\end{tabular}
    \caption{Optimization process for two variables \texttt{u} and \texttt{v} from the data set by Matsunobu et al.~\cite{MatsunobuEnsemble}. Transfer function optimization cannot find a mapping that yields similar images of both variables, confirmed also by the residual map after optimization. This hints on significant structural differences in the two compared fields.}
    \label{fig:matsunobu-uv}
\end{figure*}

\autoref{fig:necker-pearson-mi} shows another example using the data set by Necker et al., where again correlation measures are compared. This time, however, two reference points are selected, and the correlation fields for both points are subtracted to obtain a new scalar field. Once, the mutual information is used as correlation measure, and once PPMCC. The mutual information, which is computed using the estimator by Kraskov et al.~\cite{KraskovMI}, is a measure of statistical dependence from probability theory based on entropy. Thus, this measure is expected to encode fundamentally different information than PPMCC. However, it can be seen that through the optimal choice of the transfer function, the image differences between the two measures are drastically reduced. This is also shown by the residual map after optimization, which clearly indicates good agreement between the two measures in most of the domain. Again, the results are independent of the chosen least squares solver.

In a last experiment we shed light on the optimization result when two significantly different fields are compared. In this case, one can expect that the same target color in the reference voxel colors (or reference images for DiffDVR) will be related to many entries of the transfer function subject to optimization, leading to an averaged-out transfer function. We confirm this by using the longitudinal and latitudinal wind components \texttt{u} and \texttt{v} from the data set by Matsunobu et al.~\cite{MatsunobuEnsemble}.
As can be seen in \autoref{fig:matsunobu-uv}, transfer function optimization reaches its limit for these fields with significant structural differences. The resulting transfer function and DVR rendered image contain mostly smoothed out colors. The residual map indicates highly dissimilar fields, and, thus, proves to be effective in supporting a comparative analysis.

\subsection{Solver Comparison}

\definecolor{dgreen}{rgb}{0.0, 0.42, 0.24}
\newcommand{\cmark}{\textcolor{dgreen}{\ding{52}}}
\newcommand{\xmark}{\textcolor{red}{\ding{56}}}
\newcommand{\tmark}{\textcolor{orange}{$\bm{\Delta}$}}
\newcommand{\plus}{\textcolor{dgreen}{\textbf{+}}}
\newcommand{\plusplus}{\textcolor{dgreen}{\textbf{+ +}}}
\newcommand{\minus}{\textcolor{red}{\textbf{--}}}
\newcommand{\minusminus}{\textcolor{red}{\textbf{-- --}}}
\newcommand{\minusplus}{\textcolor{red}{\textbf{--}} / \textcolor{dgreen}{\textbf{+}}}
\newcommand{\plusminus}{\textcolor{dgreen}{\textbf{+}} / \textcolor{red}{\textbf{--}}}
\renewcommand{\tabularxcolumn}[1]{m{#1}}
\renewcommand{\arraystretch}{1.4}

\newcolumntype{M}{>{\centering}p{2em}}

\begin{table*}[ht]
        \setlength{\tabcolsep}{4.55pt}
	\fontsize{8pt}{8pt}\selectfont
	\centering%
	\begin{tabularx}{1.0\linewidth}
		{|X<{\centering}|c|c|c|c|c|c|c|c|}%
		\hline
		\multicolumn{1}{|>{\centering}c|}{\textbf{Solver}} 
		& \multicolumn{1}{>{\centering}c|}{\textbf{Eigen QP}} 
		& \multicolumn{1}{>{\centering}c|}{\textbf{Eigen NormEq}}
		& \multicolumn{1}{>{\centering}c|}{\textbf{Eigen CGLS}}
		& \multicolumn{1}{>{\centering}c|}{\textbf{CUDA NormEq}}
		& \multicolumn{1}{>{\centering}c|}{\textbf{CUDA CGLS}}
		& \multicolumn{1}{>{\centering}c|}{\textbf{Vulkan NormEq}}
		& \multicolumn{1}{>{\centering}c|}{\textbf{GradDesc}}
		& \multicolumn{1}{>{\centering}c|}{\textbf{DiffDVR}}
		\\ \hline
		
		Stability   & \tmark & \tmark & \cmark & \tmark & \cmark & \tmark & \cmark & \cmark \\ \hline
		Constraints & \cmark & \tmark & \tmark & \tmark & \tmark & \tmark & \tmark & \tmark \\ \hline
		Portability & \cmark & \cmark & \cmark & \xmark & \xmark & \cmark & \cmark & \cmark \\ \hline
		Misc. Norms & \xmark & \xmark & \xmark & \xmark & \xmark & \xmark & \cmark & \cmark \\ \hline
		Timings Necker \#1 & 0.59s  & 0.59s  & 0.51s  & 0.02s & 0.31s  & 0.02s  & 0.78s  & 15.20s \\ \hline
		Timings Combustion & 12.03s & 12.02s & 76.62s & 0.24s & 7.38s  & 0.34s  & 41.09s & 29.73s \\ \hline
		Timings Necker \#2 & 0.55s  & 0.54s  & 0.50s  & 0.02s & 0.25s  & 0.02s  & 0.77s  & 15.53s \\ \hline
		Timings Matsunobu  & 9.70s  & 9.53s  & 10.26s & 0.22s & 4.05s  & 0.23s  & 24.52s & 34.79s \\ \hline

	\end{tabularx}%
	\vspace{0.2cm}
	\caption{Overview over the different implemented solvers and their pros and cons. Timings for the time until convergence are given wrt.\ the optimization process in \autoref{fig:necker-pearson-kendall} for a system with an AMD Ryzen 9 3900X 12-core CPU and an NVIDIA GeForce RTX 3090 GPU. \textit{Eigen QP} stands for computing $A^T A$ with Eigen and quadratic programming via OSQP and \textit{NormEq} for solving the normal equations. Necker \#1 corresponds to \autoref{fig:necker-pearson-kendall}, Combustion to \autoref{fig:combustion}, Necker \#2 to \autoref{fig:necker-pearson-mi} and Matsunobu to \autoref{fig:matsunobu-uv}. For descriptions and sizes of these data sets, please refer to \autoref{sec:demonstration}.}
	\label{tab:comparison-solvers}
	\vspace{-0.3cm}
\end{table*}

\autoref{tab:comparison-solvers} compares the different solvers we have used and developed regarding their specific capabilities for transfer function optimization. Notably, when using voxel-based transfer function optimization, noticeable quality differences in the results could not be perceived in any of our examples. The difference between voxel- and image-based approaches are clearly visible, on the other hand. 

The most striking observation is the huge performance differences between different types of solver. Especially gradient descent and DiffDVR are significantly slower than all other alternatives, especially prohibiting the use of DiffDVR in time-critical applications or applications where new fields are instantly generated and compared.

Regarding numerical stability, solvers which do not use the normal equations have an advantage as discussed in \autoref{sec:solvers}, but are slower in general. Even though numerical stability was not a concern in any of our tests, for larger data sets this might become problematic and should be considered when selecting an appropriate solver. In all our test cases, when taking into account pivoting as mentioned in \autoref{sec:normal-eq-solvers}, the solvers based on normal equations were sufficiently numerically stable in all experiments and gave equivalent results. Despite this observation, our system supports estimating the condition number of $A^T A$, and we propose switching to other solvers, like CGLS or gradient descent, when the matrix is particularly ill-conditioned.

On the other hand, only quadratic programming can explicitly take into account the box constraints, while all other solvers need to rely on the truncation of the solution, i.e., $x_i' = \max\{\min \{ x_i, 1 \}, 0\}$. All CUDA-based solvers cannot be used on other hardware than NVIDIA GPUs, as they rely on linear algebra subroutines implemented in cuSPARSE. Notably, only gradient descent and DiffDVR can use arbitrary norms with no additional adjustments of the solvers themselves, except the modification of derivatives with respect to the used loss term. 

\begin{table}[ht]
        \setlength{\tabcolsep}{4pt}
	\fontsize{8pt}{8pt}\selectfont
	\centering%
	\begin{tabularx}{1.0\linewidth}{|X<{\centering}|m{14mm}<{\centering}|m{14mm}<{\centering}|m{14mm}<{\centering}|m{14mm}<{\centering}|}%
		\hline
		\multicolumn{1}{|>{\centering}c|}{\textbf{Solver}} 
		& \multicolumn{1}{>{\centering}c|}{\textbf{Original}} 
		& \multicolumn{1}{>{\centering}c|}{\textbf{CGLS}}
		& \multicolumn{1}{>{\centering}c|}{\textbf{DiffDVR $l2$}}
		\\ \hline
		
		RMSE            & 3.472 & 14.65 & 3.761 \\ \hline
		PSNR            & 37.31 & 24.81 & 36.62 \\ \hline
		SSIM            & 0.989 & 0.915 & 0.894 \\ \hline
		LPIPS (AlexNet) & 0.017 & 0.050 & 0.021 \\ \hline
		LPIPS (VGGNet)  & 0.037 & 0.075 & 0.044 \\ \hline

	\end{tabularx}%
	\vspace{0.2cm}
	\caption{Similarity metrics for the images in \autoref{fig:cube}. For PSNR and SSIM, higher is better. For RMSE and LPIPS, lower is better.}
	\label{tab:cube-img-metrics}
\end{table}

\begin{figure*}[t]
    \centering
\setlength{\tabcolsep}{1pt}
\begin{tabular}{cccccc}
$F_r$ (Slice) & $F_o$ (Slice) & $F_r$ (ref.) & $F_o$  & $F_o$ (opt.\ CGLS) & $F_o$ (opt.\ DiffDVR $l2$) \\
\includegraphics[width=0.16\linewidth,valign=m]{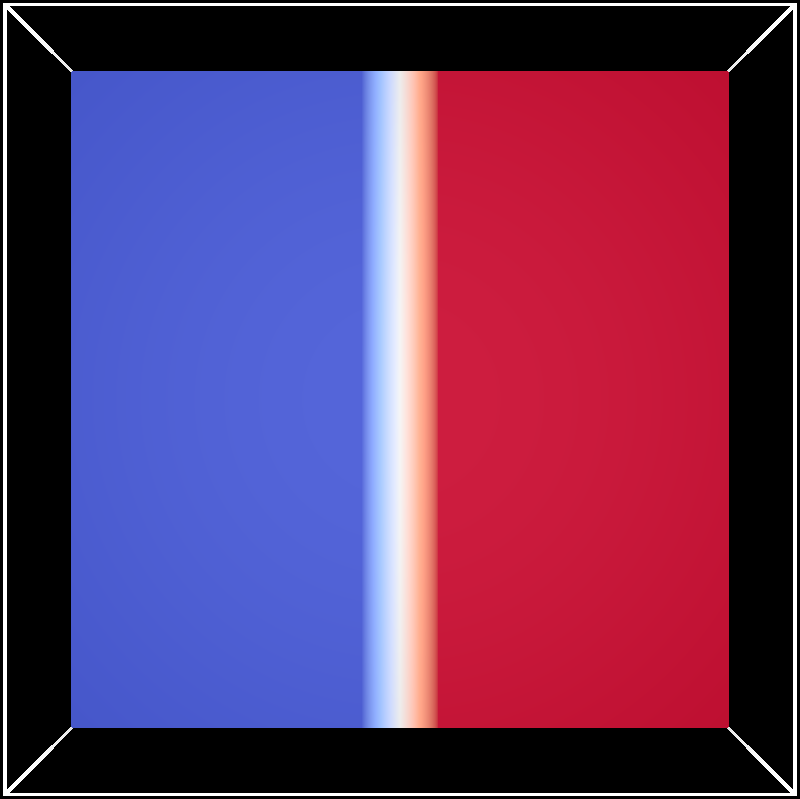} & 
\includegraphics[width=0.16\linewidth,valign=m]{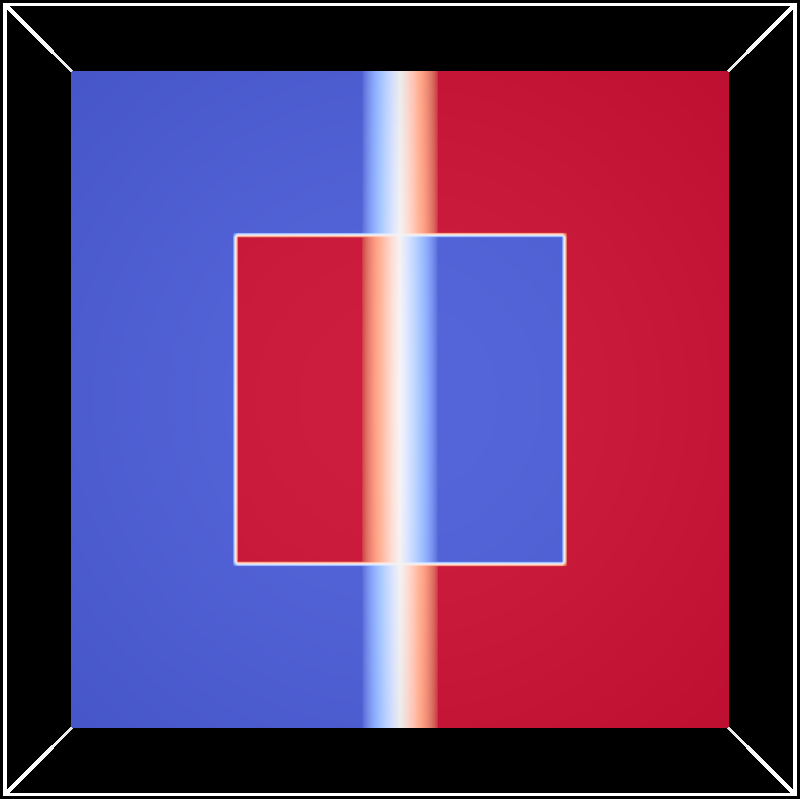} & 
\includegraphics[width=0.16\linewidth,valign=m]{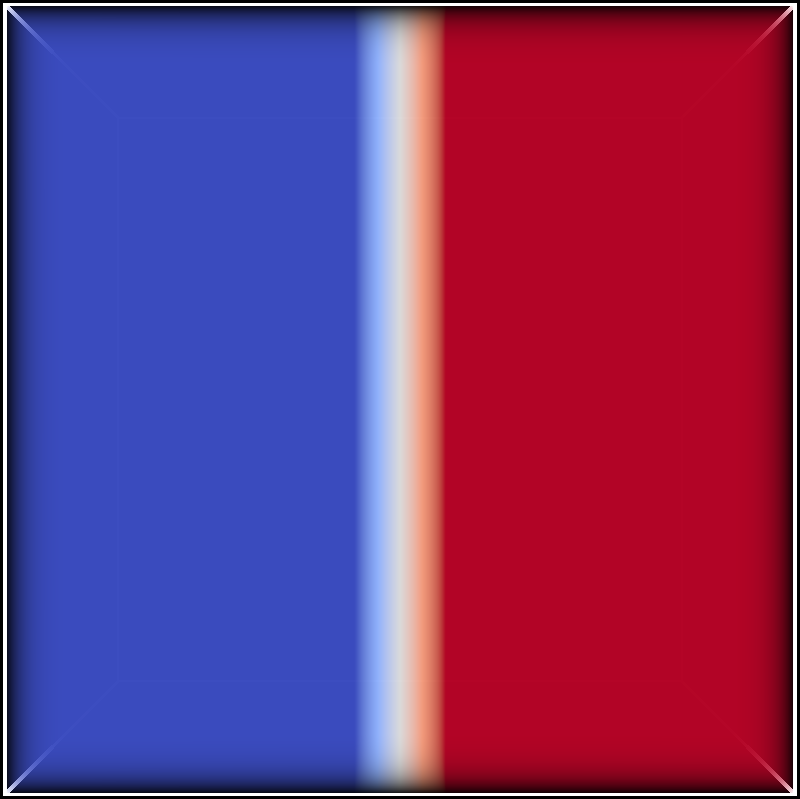} &
\includegraphics[width=0.16\linewidth,valign=m]{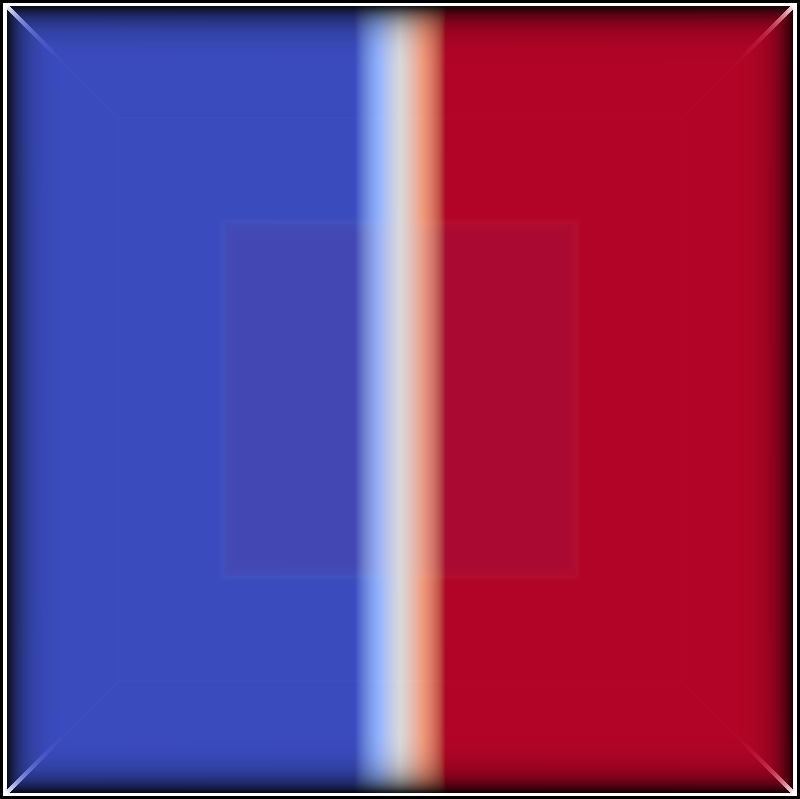} &
\includegraphics[width=0.16\linewidth,valign=m]{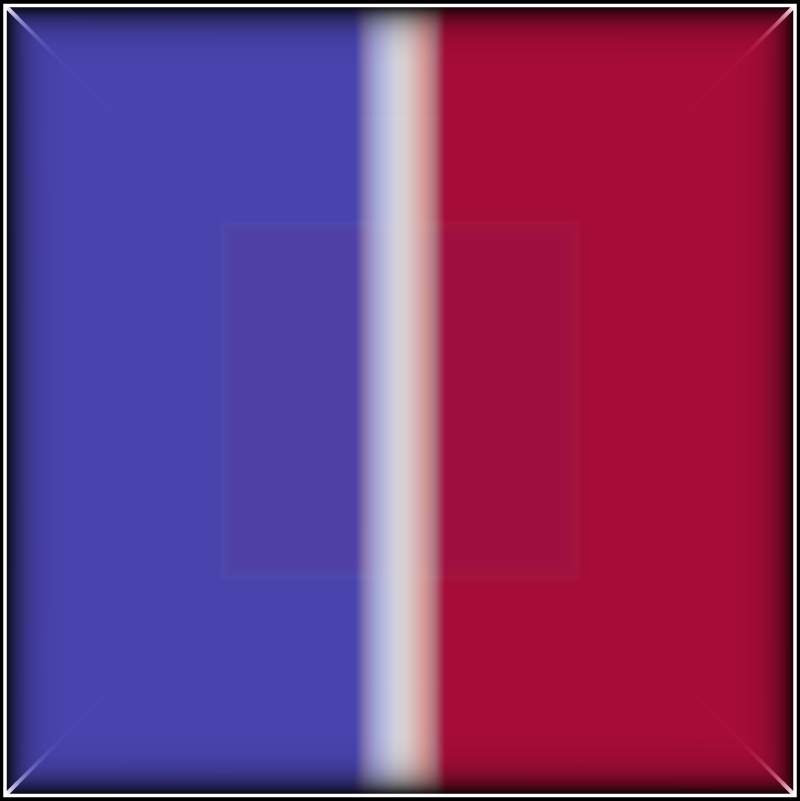} & 
\includegraphics[width=0.16\linewidth,valign=m]{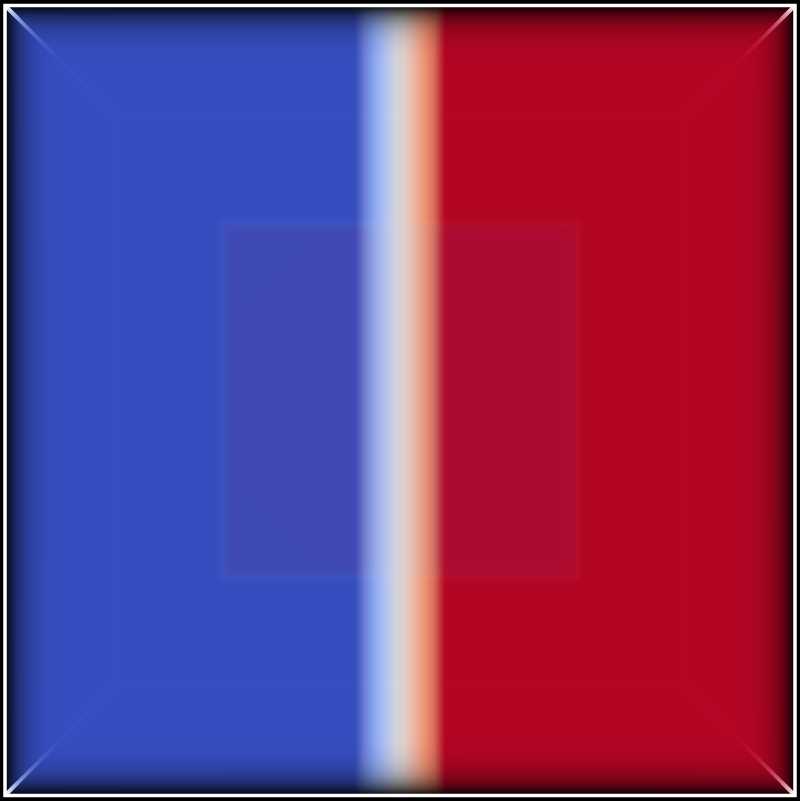} \vspace{0.1cm} \\
\end{tabular}
    \caption{Transfer function optimization for a synthetic test case. The resulting images for the least squares solvers deviate more from the reference visually since they give more emphasis to the interior of the data set.}
    \label{fig:cube}
\end{figure*}

In \autoref{fig:cube}, we further compare the capabilities of least squares solvers and DiffDVR for two fields from a synthetic data set. One scalar field, the reference field, is a volume with value $0$ on the left side of the domain and value $1$ on the right side. The second volume subject to transfer function optimization additionally contains a smaller cube in the interior with opposite value distribution. The values on the inside can hardly be seen. DiffDVR has higher gradients for more strongly visible voxels, and accordingly also hardly changes the transfer function. The voxel-based solvers, on the other hand, face the difficulty that they need to map both values $0$ and $1$ to both red and blue, and accordingly choose an averaged-out result. As a result, DiffDVR manages to get better image similarity metrics in \autoref{tab:cube-img-metrics}. 

In the end, the user needs to decide whether a solver is favoured that  more closely matches the rendered images and weighs down less visible voxels, or a solver that takes into account all values present in the data. Another disadvantage of DiffDVR is, as described by Weiss and Westermann \cite{DiffDVR}, that the problem it solves is not convex, and consequently, it might get stuck in a local minimum in the optimization process. As can be seen already in the teaser image by Weiss and Westermann \cite{DiffDVR}, DiffDVR is not even always able to perfectly reconstruct the transfer function for a data set from rendered images of itself, due to the ambiguity that slightly different transfer functions might generate almost exactly the same images.

\section{Conclusion and Future Work}

We have demonstrated that transfer functions for volume rendering can be optimized automatically so that rendered images of a volume match images of a given reference. We have introduced such optimizations for comparative volume rendering, to quickly find a transfer function that compensates non-structural differences. To visually convey structural differences, a residual map between the two pre-shaded volumes has been introduced. We have shown that the optimization can be performed efficiently using least squares solvers minimizing residuals between pre-shaded voxel colors, and compared these results to image-based optimization using differentiable direct volume rendering (DiffDVR). While DiffDVR is significantly slower than least squares approaches, it focuses the optimization on those structures contributing to the final images. As such, in the residual maps DiffDVR emphasizes more strongly those regions where structural differences are seen, also considering occlusions and attenuation due to ray-based integration. 
In our publicly available framework for automatic inter-field optimization of transfer functions we provide different solvers which can be selected and compared via an interactive user interface. 

Since in the current work an initial reference transfer function needs to be defined by the user, this restricts the structures to be conveyed to those selected by this transfer function. Thus, important regions exhibiting structural variations and differences can be overlooked. To overcome this limitation, we will investigate the integration of entropy-based loss functions to automatically determine the reference transfer function. Furthermore, we will look into approaches to optimize a transfer function by considering two volumes simultaneously in the optimization process. In this way we aim to optimize for a transfer function that reveals differences in the two volumes most effectively. Finally, we shed light on the question whether DiffDVR can be used to modify the values of one field in such a way that the image of a reference field is matched. This requires to consider additional constraints in the optimization, for instance, to maintain the histogram structure.

\section*{Acknowledgments}

Access to the convective-scale 1000 ensemble member simulation forecast by Necker et al.~\cite{Necker2020} can be requested from the authors. The turbulent combustion data set was provided by Dr.\ Jacqueline Chen from the Sandia National Lab through the SciDAC Institute for Ultra-Scale Visualization. The data set by Matsunobu et al.~\cite{MatsunobuEnsemble} is publically available.

\bibliographystyle{eg-alpha-doi} 
\bibliography{main}       


\end{document}